\documentclass[12pt,reqno,a4paper]{amsart}
\usepackage{amssymb}
\usepackage{amsmath, amsthm}
\allowdisplaybreaks[1]
\usepackage{graphicx,color}
\usepackage{wrapfig,framed,bbm}
\usepackage{microtype}
\usepackage[height=23.4cm, width=17cm, hmarginratio={1:1}]{geometry}
\usepackage{hyperref}

\theoremstyle{plain}
\newtheorem{theorem}{Theorem}
\newtheorem{prop}{Proposition}
\newtheorem{lemma}{Lemma}
\newtheorem{cor}{Corollary}
\theoremstyle{definition}
\newtheorem{defi}{Definition}
\theoremstyle{remark}
\newtheorem*{remark}{Remark}

\newcommand{\beq}{\begin{equation}}
\newcommand{\eeq}{\end{equation}}
\newcommand{\nn}{\nonumber}

\newcommand{\pal}{\partial}

\newcommand{\CC}{\mathbb{C}}

\newcommand{\F}{{\mathcal F}}

\newcommand{\bt}{{\bf t}}

\newcommand{\ZZ}{{\mathbb Z}}
\newcommand{\QQ}{{\mathbb Q}}
\newcommand{\V}{{\mathcal V}}
\newcommand{\R}{{\mathcal R}}
\newcommand{\cH}{{\mathcal H}}
\newcommand{\pf}{\noindent{\bf Proof \ }}
\newcommand{\tr}{{\rm tr}}
\newcommand{\p}{\partial}
\newcommand{\e}{\epsilon}

\newcommand{\epf}{$\quad$\hfill
\raisebox{0.11truecm}{\fbox{}}\par\vskip0.4truecm}

\def\={\; = \;}
\def\+{\, + \,}
\def\:={\; := \; }
\def\1{\mathbbm{1}}

\def\bll{\bigl\langle}
\def\brr{\bigr\rangle}

\newcommand{\indicationfootnote}{\thanks}

\begin{document}
\title{Matrix resolvent and the discrete KdV hierarchy}
\author{Boris Dubrovin$^{\dagger}$}
\indicationfootnote{$^{\dagger}$Deceased on March 19, 2019.}
\author{Di Yang}
\date{}
\maketitle
\begin{abstract}
Based on the matrix-resolvent approach, for an arbitrary solution to the discrete KdV hierarchy, 
we define the tau-function of the solution, and compare it with 
another tau-function of the solution defined via reduction of the Toda lattice hierarchy.
Explicit formulae for generating series of logarithmic derivatives of the tau-functions 
are obtained, and applications to enumeration of ribbon graphs 
 with even valencies and to certain special cubic Hodge integrals are considered. 
\end{abstract}



\setcounter{tocdepth}{1}
\tableofcontents

\section{Introduction}\label{section1}

The discrete KdV equation ({\it aka the Volterra lattice equation}) is an integrable 
 Hamiltonian equation in (1+1) dimensions, i.e. one discrete space variable and 
one continuous time variable, which extends to 
a commuting system of Hamiltonian equations, 
called the discrete KdV integrable hierarchy. This integrable hierarchy has important applications 
in algebraic geometry and symplectic geometry 
(in particular in the theory of Riemann surfaces) (see e.g.~\cite{FV}). 
Significance of the discrete KdV hierarchy was further pointed out by 
E.~Witten~\cite{Wi} in the study of 
the GUE partition function with even couplings --- the ``{\it matrix gravity}", and was recently
 addressed also in the study of the special cubic 
 Hodge partition function \cite{DuY2,DLYZ1,DLYZ2} --- 
 the {\it topological gravity} in the sense of~\cite{DuY2,DLYZ2}. 
 The explicit relationship between the two gravities, 
 called the Hodge--GUE correspondence, 
 has been established in~\cite{DuY2,DLYZ2}. In this paper, 
 by using the matrix-resolvent (MR) approach recently introduced and 
 developed in~\cite{BDY1,BDY20,BDY2,DuY1,DYZ2} we study the tau-structure for the 
discrete KdV hierarchy, and apply it to studying the above mentioned enumerative problems.

\medskip

\subsection{The discrete KdV hierarchy}
Let $P(n)$ be the following difference operator
\beq
P(n) \:= \Lambda + w_n\, \Lambda^{-1} \, , 
\eeq
where $\Lambda$ denotes the shift operator $\Lambda: f_n \mapsto f_{n+1}$.  Introduce
\beq
A_\ell \:= \bigl(P^{\ell+1}\bigr)_+\,, \quad \ell\geq 0\,.
\eeq
Here, for an operator~$Q$ of the form $Q = \sum_{k \in \mathbb{Z}} Q_k \, \Lambda^k$, 
the positive part $Q_+ := \sum_{k\geq 0} Q_k \, \Lambda^k$. 
The discrete KdV hierarchy is defined as the following system of commuting flows:
\beq \label{defred}
\frac{\p P}{\p s_j} \=  \bigl[ A_{2j-1} \,,\, P \bigr]\,,\qquad j\geq 1\,.
\eeq
For example, the $s_1$-flow reads
\beq\label{firstflow}
\frac{\p w_n}{\p s_1} \=  w_n \, (w_{n+1} - w_{n-1}) \,, 
\eeq
which is the {\it discrete KdV equation}. The commutativity implies that equations \eqref{defred} for all $j\geq 1$ can be solved together, 
yielding solutions of the form $w_n=w_n({\bf s})$,~${\bf s}:=(s_1,s_2,s_3,\dots)$. 

Let us introduce 
\beq\label{definitionL}
L \:= P^2 \= \Lambda^2 \+ w_{n+1} + w_n \+ w_n \, w_{n-1} \, \Lambda^{-2}\,.
\eeq
Then $A_{2j-1} = \bigl(P^{2j}\bigr)_+ = \bigl(L^j\bigr)_+$\,.  
\begin{lemma}\label{equivdkdv}
The discrete KdV hierarchy~\eqref{defred} can be equivalently written as
\beq \label{dkdv}
\frac{\p L}{\p s_j} \=  \bigl[ A_{2j-1} \,,\, L \bigr] \,,\qquad j\geq 1\,.
\eeq
\end{lemma}
\noindent 
The proof will be given in Section~\ref{sect23}.
For the particular case $j=1$, we have
\begin{align}
& \frac{\p (w_{n+1} + w_n)}{\p s_1} \= w_{n+2} \, w_{n+1} - w_n \, w_{n-1}\,,  \label{1n1}\\
& \frac{\p (w_n \, w_{n-1})}{\p s_1} \=  (w_{n+1} + w_n - w_{n-1} - w_{n-2}) \, w_n \, w_{n-1}\label{1n2}.
\end{align}
It can be shown that equations \eqref{1n1}--\eqref{1n2} are equivalent to equation~\eqref{firstflow}; the details 
for this equivalence are in Section~\ref{section2point2}.

Observe that equations~\eqref{dkdv} are the compatibility conditions of the following scalar Lax pairs:
\begin{align} 
& L \psi_n \= \lambda \, \psi_n\,, \quad \mbox{i.e.} \quad \psi_{n+2}  + (w_{n+1} + w_n -\lambda) \, \psi_n + w_n \, w_{n-1} \, \psi_{n-2}  \= 0\,, \label{scalarl} \\
& \frac{\p \psi_n}{\p s_j} \= A_{2j-1} \, \psi \,.  \label{scalara}
\end{align}
We want to write the spectral problem \eqref{scalarl} into a matrix form.
The scalar Lax operator $L$, defined in \eqref{definitionL}, could be viewed as a reduction of 
$$
\widetilde L \= \Lambda^2 + a_1(n) \, \Lambda + a_2(n) + a_3(n) \, \Lambda^{-1} + a_4(n) \, \Lambda^{-2}\,,
$$
which is the Lax operator of a bigraded Toda hierarchy.
However, observe that $L$ contains $\Lambda^{{\rm even}}$ only (with ${\rm even} =-2,0,2$).
So, instead of considering a $4\times4$ matrix-valued Lax operator, a $2\times 2$ matrix-valued operator will be sufficient. 
Indeed, introduce 
\beq
{\mathcal L} \:=  \begin{pmatrix} \Lambda^2  & 0 \\ 0 & \Lambda^2 \\  \end{pmatrix} \+ U_n \, , 
\qquad U_n \:= \begin{pmatrix}  w_{n+1} + w_n -\lambda & w_n \, w_{n-1}  \\ -1 & 0\\  \end{pmatrix}\,.
\eeq
Then the spectral problem \eqref{scalarl} reads
\beq\label{matrixl}
\mathcal {L} \, \begin{pmatrix} \psi_n   \\ \psi_{n-2} \\  \end{pmatrix} \= 0\,.
\eeq

\subsection{The MR approach to tau-functions} In this subsection, 
we apply the MR approach to study further some basics in the theory of the 
discrete KdV hierarchy (in particular about tau-function), and will arrive at a formula for computing logarithm of the tau-function.
Denote by~$\ZZ[{\bf w}]$ the ring of polynomials with integer coefficients in the variables ${\bf w}:=(w_{n+i})|_{i\in \ZZ}$. 
\begin{defi}
An element 
$R_n\in {\rm Mat} \bigl(2, \ZZ[{\bf w}](\hskip -0.05truecm(\lambda^{-1})\hskip -0.05truecm) \bigr)$ 
is called a {\bf matrix resolvent} (MR) of~${\mathcal L}$, if
\beq
R_{n+2} \, U_n \,-\, U_n \, R_n \= 0\,.
\eeq
\end{defi}

\begin{defi} \label{rdefi} 
The basic (matrix) resolvent~$R_n$ is defined as the MR of~${\mathcal L}$ satisfying  
\begin{align}
& R_n \=  \begin{pmatrix} 1  & 0 \\ 0 & 0 \\  \end{pmatrix}  + \mathcal{O}\bigl(\lambda^{-1}\bigr) \,, \label{rini}\\ 
& \tr \, R_n \= 1\,, \quad {\rm det} \, R_ n \= 0\,.
\end{align} 
\end{defi}
\noindent The basic resolvent $R_n$ exists and is unique. See in Section~\ref{MRsection} for the proof.  Write
\beq
R_n (\lambda) \= \begin{pmatrix}  1+ \alpha_n(\lambda)  & \beta_n(\lambda) \\ \gamma_n(\lambda) & -\alpha_n(\lambda) \\  \end{pmatrix} \,.
\eeq
Then Definition \ref{rdefi} for $R_n(\lambda)$ is equivalent to the following set of equations
\begin{align}
& \beta_n \= - w_n \, w_{n-1} \, \gamma_{n+2} \label{rrr1} \\
& \alpha_{n+2}+ \alpha_n + 1 \= (\lambda-w_{n+1}- w_n) \, \gamma_{n+2}  \label{rrr2} \\
& (\lambda-w_{n+1}- w_n) (\alpha_n- \alpha_{n+2}) \= w_n \, w_{n-1} \, \gamma_n - w_{n+2} \, w_{n+1} \, \gamma_{n+4} \label{rrr3}\\
& \alpha_n + \alpha_n^2 + \beta_n \, \gamma_n=0 \label{rrr4}
\end{align}
together with equation \eqref{rini}.  These equations give recursive relations and initial values for the coefficients 
of $\alpha_n, \, \beta_n,\, \gamma_n$ (see \eqref{mr1}--\eqref{mr3} below),
which will be called the {\it MR recursive relations}. 

\begin{lemma} \label{fromMRtotau} 
For an arbitrary solution~$w_n({\bf s})$ to the discrete KdV hierarchy, let~$R_n(\lambda)$ 
denote the basic resolvent of~$\mathcal{L}$ evaluated at $w_n = w_n({\bf s})$. 
There exists a function $\tau_n^{\textsc{\tiny\rm dKdV}}({\bf s})$  satisfying 
\begin{align}
& \sum_{i,j\geq 1} \frac{\p^2 \log \tau_n^{\textsc{\tiny\rm dKdV}} ({\bf s})}{\p s_i \p s_j} \lambda^{-i-1} \mu^{-j-1} \= \frac{\tr \, \bigl(R_n(\lambda) R_n(\mu)\bigr) -1 }{(\lambda-\mu)^2 } \,, \label{d1}\\
&  \frac1{\lambda} + \sum_{i\geq 1} \frac{1}{\lambda^{i+1}} \frac{\p }{\p s_i} \log \frac{\tau^{\textsc{\tiny\rm dKdV}}_{n+2}}{\tau^{\textsc{\tiny\rm dKdV}}_{n}} \= \bigl[R_{n+2}(\lambda)\bigr]_{21}\,, \label{d2}\\
& \frac{\tau^{\textsc{\tiny\rm dKdV}}_{n+2} \, \tau^{\textsc{\tiny\rm dKdV}}_{n-1}}{ \tau^{\textsc{\tiny\rm dKdV}}_{n+1} \, \tau^{\textsc{\tiny\rm dKdV}}_n} \= w_n\,. \label{d3}
\end{align} 
Moreover, the function $\tau_n^{\textsc{\tiny\rm dKdV}}({\bf s})$ is 
uniquely determined by~$w_n({\bf s})$ up to a factor of the form
$$
e^{\alpha n + \beta_0  +\sum_{k\geq 1} \beta_k s_k}\,,
$$
where $\alpha,\beta_0,\beta_1,\beta_2,\cdots$ are arbitrary constants that are independent of $n, {\bf s}$.
\end{lemma}
\noindent We call $\tau_n^{\textsc{\tiny\rm dKdV}}({\bf s})$ the {\it tau-function} 
of the solution~$w_n=w_n({\bf s})$ to the discrete KdV hierarchy.

The matrix-resolvent method then allows to compute  
logarithmic derivatives of $\tau_n^{\textsc{\tiny\rm dKdV}}({\bf s})$, which is achieved via the following proposition.
\begin{prop} \label{main1}
For any $k\geq 3$, the generating series of the 
$k_{th}$-order logarithmic derivatives of $\tau_n^{\textsc{\tiny\rm dKdV}}({\bf s})$
 has the following expression:
\beq
\sum_{j_1,\dots,j_k=1}^\infty \frac1{\lambda_1^{j_1+1} \cdots  \lambda_k^{j_k+1} }
\frac{\p^k \log \tau_n^{\textsc{\tiny\rm dKdV}}({\bf s})}{\p s_{j_1} \dots \p s_{j_k}} \= -\frac1{k} 
\sum_{\sigma \in S_k}  \frac{ \tr \, \bigl(R_n(\lambda_{\sigma_1}) \cdots R_n(\lambda_{\sigma_k})\bigr)} 
{\prod_{i=1}^k (\lambda_{\sigma_i}-\lambda_{\sigma_i+1})} \,, 
\eeq
where it is understood that $\sigma_{k+1}=\sigma_1$.
\end{prop}
\noindent The proof of this proposition is in Section~\ref{Generatingsubsection}. 

\subsection{The factorization formula} 
In~\cite{DuY1} we gave the definition of tau-function for the Toda lattice using the MR approach. 
Observe that the discrete KdV hierarchy~\eqref{defred} is a reduction of the Toda lattice hierarchy.
Therefore, for the arbitrary solution~$w_n({\bf s})$ to the discrete KdV, 
we can also associate another tau-function~$\tau_n({\bf s})$ of the solution~$w_n({\bf s})$
obtained via the reduction (see Section \ref{reductionsection} for the precise definition). In particular, 
this tau-function satisfies that 
$$
w_n({\bf s}) \= \frac{\tau_{n+1}({\bf s}) \, \tau_{n-1} ({\bf s})}{\tau_n^2({\bf s})}  \,. 
$$
It turns out that the~$\tau_n({\bf s})$ factorizes into a product of two as given by the following theorem.
\begin{theorem} \label{maintheorem} 
There exist constants $\alpha,\beta_0,\beta_1,\beta_2,\cdots$ such that 
\beq \label{identitymaintt}
\tau_n({\bf s}) \= e^{\alpha n + \beta_0  + \sum_{k\geq 1} \beta_k s_k} \, 
\tau_n^{\textsc{\tiny\rm dKdV}}({\bf s}) \, \tau_{n+1}^{\textsc{\tiny\rm dKdV}}({\bf s}) \,.
\eeq
\end{theorem}
\noindent The proof of this theorem is in Section~\ref{SectionMaintheorem}.

\begin{remark}
Identity~\eqref{identitymaintt} echoes an identity between Hankel determinants. Indeed, let $d\mu(\lambda)$ be a measure 
with even moments on~$\mathbb{R}$. Denote $\mu_j=\int \lambda^j d\mu(\lambda)$, $j\geq 0$. ($\mu_{\rm odd}=0$.)
We know that 
\beq\label{factorhankel}
\det \bigl( \mu_{i+j-2} \bigr)_{i,j=1}^{n}  \=\det \bigl( \mu_{2i+2j-2} \bigr)_{i,j=1}^{[n/2]}   \det \bigl( \mu_{2i+2j-4} \bigr)_{i,j=1}^{[(n+1)/2]}  \,.
\eeq
If we deform the measure $d\mu(\lambda)$ to be $d\mu(\lambda;\bt)=e^{-\sum_{j\geq1} t_{j-1} \lambda^j} d\mu(\lambda)$, 
then the LHS$\cdot (2\pi)^{-n}$ becomes a Toda tau-function (cf. the formula (3.9) of~\cite{Deift} 
and the references therein; cf. also~\cite{Deift, mehta, DuY1}; the $(2\pi)^{-n}$ is a normalization 
factor for convenience that does not affect the fact that the LHS is already a Toda tau-function). If all the 
even Toda times are zero, then the $\bt$-deformed measure remains even and the factorization~\eqref{factorhankel} holds identically in 
$\bt =(0,s_1,0,s_2,\cdots)$. 
Moreover, note that the RHS of~\eqref{factorhankel} with deformation consists of two determinants which can be identified with 
the Hankel determinants 
associated with certain ${\bf s}$-deformed measures on $\mathbb{R}_+$, where ${\bf s}=(s_1,s_2,\cdots)$.
Then to see~\eqref{identitymaintt} from~\eqref{factorhankel}, at least for special cases, 
one needs to further show that each of the {\it two} determinants is a tau-function for the discrete KdV hierarchy. 
The more precise statement for a special case together with the detailed proofs can be seen from 
the recent arXiv preprint by Massimo Gisonni, Giulio Ruzza and Tamara Grava~\cite{GGR} 
regarding Laguerre Unitary Ensemble (LUE) with the consideration of 
the parameters $\alpha=-1/2$ and $\alpha=1/2$, {\it respectively} in the notations of~\cite{GGR} (cf. also~\cite{CGM,CDOC}).
\end{remark}

The next corollary follows from Proposition~\ref{main1} and Theorem~\ref{maintheorem}.

\begin{cor} \label{cor1} Fix $k\geq 2$. Let $w_n=w_n({\bf s})$ be an arbitrary solution to the discrete KdV hierarchy, 
and $\tau_n$ the tau-function reduced from
the Toda lattice hierarchy of~$w_n({\bf s})$. The following formula holds true:
\begin{align}
\sum_{j_1,\dots,j_k=1}^\infty \frac{\frac{\p^k \log \tau_n({\bf s})}{\p s_{j_1} \dots \p s_{j_k}}}{\lambda_1^{j_1+1} \cdots  \lambda_k^{j_k+1} }
 \= & -\frac1{k}  
\sum_{\sigma \in S_k}  \frac{ (1+\Lambda) \,  \tr \bigl[R_n(\lambda_{\sigma_1}) \cdots R_n(\lambda_{\sigma_k}) \bigr]}
{\prod_{i=1}^k (\lambda_{\sigma_i}-\lambda_{\sigma_i+1})} - \frac{2\, \delta_{k,2}}{(\lambda_1-\lambda_2)^2} \,.
\end{align}
\end{cor}

In practice, the two tau-functions $\tau_n$ and~$\tau_n^{\textsc{\tiny\rm dKdV}}$ 
of some solution for the discrete KdV hierarchy may both have  
geometric/enumerative meanings; this is the case for the Hodge--GUE (see below). 

\begin{remark}
As we shall see from Section~\ref{reductionsection} that 
the above mentioned reduction {\it does not} mean 
that $v_n^{\textsc{\tiny\rm Toda}}$, $w_n^{\textsc{\tiny\rm Toda}}$ (see Section~\ref{reductionsection})  are independent 
of the even Toda times $t_0,t_2$, $\cdots$. 
The reduction means the $v_n^{\textsc{\tiny\rm Toda}}(0,t_1,0,t_3,\cdots)\equiv 0$; 
but the usage of the {\it MR of Toda} in the way of~\cite{DuY1} would compute also 
the correlators containing the correspondence to $t_0,t_2$, $\cdots$. The introductions 
of the {\it MR of the discrete KdV hierarchy} and 
of the operator $1+\Lambda$ are essential that surprisingly solve the problem in a simple form.
\end{remark}

%
%

\subsection{Application} 
We will first apply Corollary~\ref{cor1} to 
some counting problem. Then 
by using the Hodge--GUE correspondence~\cite{DLYZ2,DuY2} 
we compute some combinations of Hodge integrals. 

\noindent {\it I. Enumeration of ribbon graphs with even valencies.}
Enumeration of ribbon graphs is closely 
related to the random matrix theory \cite{BIZ, HZ, KKN, DF, mehta, CE}: e.g. to the Gaussian Unitary Ensembles~(GUE) correlators; 
the partition function with coupling constants in a random matrix theory is often a tau-function of some integrable system. 
Given $k\geq 1$ and $j_1,\dots, j_k\geq1$, denote 
\begin{align} 
& \bll\tr \, M^{2j_1} \, \cdots \, \tr \, M^{2j_k} \brr_c \:= k! \,\sum_{0\leq g \leq \frac{|j|}2-\frac k2+\frac12}  \, n^{2-2g-k+|j|}  \, a_g(2j_1,\dots,2j_k) \, ,  \\
& a_g(2j_1, \dots, 2j_k) \:=  \sum_{\Gamma} \frac1{\#\, {\rm Sym} \, \Gamma} \, . \label{gue-ag}
\end{align}
Here, $|j|=j_1 + \dots + j_k$, and $\sum_{\Gamma}$ denotes summation over 
 connected ribbon graphs $\Gamma$ with labelled half edges and unlabelled vertices of genus~$g$ 
with $k$~vertices of valencies~$2j_1, \dots, 2j_k$, and $\#\,{\rm Sym}\, \Gamma$ is the order of the symmetry 
group of $\Gamma$ generated by permuting the vertices.\footnote{The number $a_g(2j_1,\dots,2j_k)$ has the alternative expression
$a_g(2j_1,\dots,2j_k) =  \sum_G  \frac{\prod_{\ell=1}^k (2 j_\ell)}{\#\, {\rm Sym} \, G}$,
where $\sum_G$ denotes summation over connected ribbon graphs~$G$ with 
unlabelled half-edges and unlabelled vertices of genus~$g$ with $k$~vertices of valencies~$2j_1, \dots, 2j_k$.} 
The notation $\bll\tr \, M^{2j_1} \, \cdots \, \tr \, M^{2j_k} \brr_c$ is borrowed from the literature of random matrices,  
where it is often 
called a connected Gaussian Unitary Ensemble (GUE) correlator. 
For every $k\geq 1$, denote 
\beq
E_k(n;\lambda_1,\dots,\lambda_k) \:= 
\sum_{j_1,\dots,j_k=1}^\infty \frac{ \bigl\langle \tr \,M^{2j_1} \cdots \tr \, M^{2j_k} \bigr\rangle_c}{\lambda_1^{j_1+1} \cdots \lambda_k^{j_k+1}}  \, .
\eeq
\begin{defi}\label{defRAB}
Define a $2\times 2$ matrix-valued series $R_n(\lambda)\in {\rm Mat}\bigl(2, \mathbb Z[n] [[\lambda^{-1}]] \bigr)$ by
\begin{align}
\label{ram}
 R_n(\lambda) \:= 
 \begin{pmatrix} 1 & 0 \\ 0 & 0\end{pmatrix} \+ 
{\tiny 
\sum_{j=0}^\infty \frac{(2j-1)!!}{\lambda^{j+1}} 
\begin{pmatrix}  
(2j+1) A_{n,j} - (n-1)B_{n,j} &    (n-n^2)\,B_{n+2,j}\\ \\
B_{n,j} & (n-1) B_{n,j} - (2j+1) A_{n,j}  
\end{pmatrix} }
\end{align}
with
\begin{align} 
& A_{n,j} \:= (n-1)\, {}_2F_1(-j,2-n; 2; 2) \, ,
\label{ank} \\
& B_{n,j} \:= (n-1) \, {}_2F_1(1-j,2-n; 2; 2) + (n-2) \, {}_2F_1(1-j,3-n; 2; 2) \, .
\label{bnk}
\end{align}

\end{defi}

\begin{theorem} \label{thm3} The following formulae hold true:
\begin{align}
& E_1(n;\lambda) \= n \, \sum_{j\geq 1} \frac{(2j-1)!!}{\lambda^{2j+1}} \Bigl( {}_2F_1(-j,-n; 2; 2) \,-\, j \, {}_2F_1(1-j, 1-n; 3; 2)\Bigr) \,, \label{onepoint} \\
& E_2(n; \lambda_1,\lambda_2) \= \frac{(1+\Lambda) \, \bigl[\tr\, \bigl(R_n(\lambda_1) R_n(\lambda_2)\bigr)\bigr]}{(\lambda_1-\lambda_2)^2} -\frac2{(\lambda_1-\lambda_2)^2} \,, 
\label{twopoint} \\
& E_k(n; \lambda_1, \dots, \lambda_k) \= 
-\frac1{k}\sum_{\sigma\in S_k} \frac{ (1+\Lambda) \, \bigl[\tr \,\bigl( R_n(\lambda_{\sigma_1}) \cdots  R_n(\lambda_{\sigma_k})\bigr)\bigr]}
{\prod_{\ell=1}^k (\lambda_{\sigma_\ell}-\lambda_{\sigma_\ell+1})}   \quad (k\geq 3)\,,   \label{kpoint}
\end{align}
where $R_n(\lambda)$ is defined in Definition~\ref{defRAB}, and it is understood that $\sigma_{k+1}=\sigma_1$. 
\end{theorem}
\noindent In the above formulae
$$
{}_2F_1(a,b; c; z) \= \sum_{j=0}^\infty \frac{(a)_j (b)_j}{(c)_j} \frac{z^j}{j!} \= 1 \+ \frac{a\, b}{c} \frac{z}{1!} \+ \frac{a(a+1)\, b(b+1)}{c(c+1)}\frac{z^2}{2!} \+ \cdots
$$
is the Gauss hypergeometric function. Recall that it truncates to a polynomial if $a$ or $b$ are non-positive integers. In particular, 
$$
n \, {}_2F_1(-j,1-n; 2; 2) \=  \sum_{i=0}^j 2^i \, \binom{j}{i} \binom{n}{i+1} \,. 
$$
\noindent The proof of Theorem~\ref{thm3} is in Section~\ref{appsection}.

\noindent {\it II. Combinations of certain special cubic Hodge integrals.} 
The particular solution to the discrete KdV hierarchy considered here will be actually 
the same as in I. 
Denote by $\overline{\mathcal{M}}_{g,k}$ the Deligne--Mumford moduli space of stable algebraic curves of genus $g$ with 
$k$ distinct marked points, by $\mathcal{L}_i$ the $i_{\rm th}$ tautological line bundle on $\overline{\mathcal{M}}_{g,k}$, 
and $\mathbb{E}_{g,k}$ the Hodge bundle. Denote
\begin{align}
& \psi_i \:= c_1(\mathcal{L}_i)\,, \quad  i=1,\dots,k \, ,\nn\\
& \lambda_j \:= c_j(\mathbb{E}_{g,k}) \, , \quad j=0,\dots,g \, . \nn
\end{align}
The Hodge integrals are some rational numbers defined by
$$
\int_{\overline{\mathcal{M}}_{g,k}} \psi_1^{i_1}\cdots \psi_k^{i_k} \, \lambda_1^{j_1} \cdots \lambda_g^{j_g} 
\; =: \; \bll \, \lambda_1^{j_1} \cdots \lambda_g^{j_g}\, \tau_{i_1}\cdots \tau_{i_k} \, \brr_{g,k}\,, \qquad i_1,\dots,i_k,\, j_1,\dots,j_g \geq 0\,. 
$$
These numbers are zero unless the degree-dimension matching is satisfied
\beq
3g\,-\,3 \+k \= \sum_{\ell=1}^k i_\ell \+ \sum_{\ell=1}^g \ell \, j_\ell \, .  \label{ddmatching}
\eeq
We are particularly interested in the following {\it special cubic Hodge integrals}:
$$
\bll \, \Omega_g\, \tau_{i_1}\cdots \tau_{i_k} \, \brr_{g,k} \,, \qquad {\rm with} \quad \Omega_g:= \Lambda_g(-1)\,\Lambda_g(-1)\,\Lambda_g\bigl(\tfrac12\bigr)\,, 
$$
where $\Lambda_g(z):= \sum_{j=0}^g \lambda_j \, z^j$ 
denotes the Chern polynomial of the Hodge bundle~$\mathbb{E}_{g,k}$. 
Significance of these Hodge integrals is manifested by the 
Gopakumar--Mari\~no--Vafa conjecture \cite{GV,MV} regarding the 
Chern--Simons/string duality; see e.g.~\cite{OP} and the references therein. 

\noindent {\it Notations.} $\mathbb{Y}$ denotes the set of partitions. 
For a partition~$\lambda$, denote by~$\ell(\lambda)$ the 
length of~$\lambda$ and by~$|\lambda|$ the weight of~$\lambda$. 
Denote $m(\lambda):=\prod_{i=1}^\infty m_i(\lambda)$ 
with~$m_i(\lambda)$ being the multiplicity of~$i$ in~$\lambda$.

\begin{defi} For given $g,k\geq 0$ and an arbitrary set of integers $i_1,\dots,i_k\geq 0$, define 
\beq\label{defhgi}
H_{g,i_1,\dots,i_k} \= 2^{g-1}  \sum_{\lambda\in \mathbb{Y}} \frac{(-1)^{\ell(\lambda)}}{m(\lambda)! } 
\bll \Omega_g \, \tau_{\lambda+1} \,\tau_I  \brr_{g, \, \ell(\lambda)+k} \,, 
\eeq
where $|i|:= i_1+\dots + i_k$, $\tau_I:=  \tau_{i_1} \cdots \tau_{i_k}$, and $\tau_{\lambda+1}:= \tau_{\lambda_1+1}\cdots \tau_{\lambda_{\ell(\lambda)}+1}$. 
\end{defi}
\noindent It should be noted that according to \eqref{ddmatching}, ``$\sum_{\lambda\in \mathbb{Y}}$" in \eqref{defhgi} is a finite sum.

The following lemma will be proved in Section~\ref{section52}.
\begin{lemma}\label{van1}
The number $H_{g,i_1,\dots,i_k}$ vanishes unless $|i|\leq 3g-3+k$.
\end{lemma}

\begin{cor} \label{app_hodge}   The numbers $H_{g,i_1,\dots,i_k}$ satisfy

i)  For $k=0$, 
\begin{align}
H_{g,\emptyset}   \=  \left\{\begin{array} {cl}
0 \,,  &  g\=0,1\,, \\
\frac{1}{4g(2g-1)(2g-2)}\sum_{g_1=0}^g (2g_1-1)\binom{2g}{2g_1} \frac{E_{2g-2g_1} B_{2g_1}}{2^{2g-2g_1}} \,, & g\geq 2\,.\\
\end{array} \right. \nn
\end{align}

ii)  For $k=1$, $\forall\, j\geq 1$, 
\begin{align}
&   \binom{2j}{j} \,   \sum_{g\geq 0} \e^{2g-1}  
\sum_{0\leq i\leq 3g-3+k}   j^{i+1}  H_{g,i}  \+  \frac{1}{2\e} \frac{1}{1+j}\binom{2j}{j} \nn\\
& \qquad\quad \=  
\e^j \biggl[\frac{(2j+1)!!}{2j} A_{\frac12+\frac1\e,\,j} \+ \frac{(2j-1)!!}{2j} \Bigl(\frac12 - \frac1\e\Bigr)B_{\frac12+\frac1\e,\,j} \biggr] \,, \label{kequals1}
\end{align}
where $A_{n,j}$ and $B_{n,j}$ are defined in~\eqref{ank}--\eqref{bnk}.

iii)  For $k\geq 2$,
\begin{align}
& \e^k \,  \sum_{j_1,\dots,j_k\geq 1} 
\frac{\prod_{r=1}^k \binom{2j_r}{j_r}} {\lambda_1^{j_1+1} \cdots \lambda_k^{j_k+1}}   \sum_{g\geq 0} \e^{2g-2}  
\sum_{i_1,\dots,i_k\geq 0 \atop |i|\leq 3g-3+k}    \prod_{r=1}^k j_\ell^{i_r+1}  H_{g,i_1,\dots,i_k}  \nn\\
& = \;  -\frac1{k}\sum_{\sigma\in S_k} \frac{ \tr \,\Bigl[ R_{\frac12+\frac1\e}\bigl(\frac{\lambda_{\sigma_1}}\e\bigr) \cdots  R_{\frac12+\frac1\e}\bigl(\frac{\lambda_{\sigma_k}}\e\bigr)\Bigr]}
{\prod_{\ell=1}^k (\lambda_{\sigma_\ell}-\lambda_{\sigma_\ell+1})} -  \frac{\delta_{k,2}}{(\lambda_1-\lambda_2)^2}  \;-\; \delta_{k,2} \, \sum_{j_1,j_2\geq 1} \frac{j_1 \, j_2}{j_1+j_2}  \frac{\binom{2j_1}{j_1} \binom{2j_2}{j_2}}{\lambda_1^{j_1+1} \lambda_2^{j_2+1}} \,,  \label{kgeq2}
\end{align}
where $R_n(\lambda)$ is defined as in~\eqref{ram}.
\end{cor}
\noindent The proof, using the Hodge--GUE correspondence and Theorem~\ref{thm3}, will be given in Section~\ref{section52}.  
We note that  the sum $\sum_{i_1,\dots,i_k\geq 0 \atop |i|\leq 3g-3+k}$ appearing in the LHS of~\eqref{kequals1}, \eqref{kgeq2}
 has the following alternative expression, which can be deduced from Appendix~\ref{appendixa}:
\begin{align}
& \sum_{i_1,\dots,i_k\geq 0 \atop |i|\leq 3g-3+k}    \prod_{r=1}^k j_r^{i_r+1}  H_{g,i_1,\dots,i_k} \= 
 \sum_{q\geq k}  \frac{1}{(q-k)!}
\int_{\overline{\mathcal{M} }_{g,q}}  \Omega_{g,q}   
\prod_{m={k+1}}^q  \biggl(-  \frac{\psi_m^2}{1-\psi_m} \biggr)  
  \prod_{m=1}^{k}\frac{j_m}{1- j_m\psi_{m}} \,. \nn
\end{align}

\noindent {\bf Organization of the paper.} 
In Section~\ref{sect23}, we derive several useful formulae. 
In Section~\ref{MRsection}, we study MR, and use it to describe the discrete KdV flows and 
the tau-structure.  Section~\ref{SectionMaintheorem} is devoted to the proof of Theorem~\ref{maintheorem}. 
Proofs of Theorem~\ref{thm3} and Corollary~\ref{app_hodge} are in Section~\ref{appsection}. 

\noindent {\bf Acknowledgements.}  
We would like to thank the anonymous referee for valuable 
suggestions and constructive comments that 
improve a lot the presentation of the paper. 
One of the authors D.Y. is grateful to Youjin Zhang and 
Don Zagier for their advising, and to   
Giulio Ruzza for helpful discussions. Part of the work of~D.Y. 
was done when he was a post-doc at MPIM, Bonn; 
he thanks MPIM for excellent working conditions and financial supports.

\section{Basic formulation} \label{sect23}

In this section we will do some preparations for the later sections by reviewing the basics of the theory of the discrete KdV hierarchy.

\subsection{Some useful identities} 
Recall that $P(n) \:= \Lambda + w_n\, \Lambda^{-1}$, $L=P^2$.  
Denote 
\begin{align}
& P(n)^{\ell+1} \; =: \; \sum_{k\in\mathbb{Z}} A_{\ell, k}(n)  \, \Lambda^k\,, \qquad \ell\geq -1\,, \label{defalk}\\
& L(n)^j \; =: \; \sum_{k\in\mathbb{Z}} m_{j,k}(n) \, \Lambda^k\,,\qquad j\geq 0 \, , \label{defmjk} 
\end{align}
where the coefficients $A_{\ell,k}(n)$ and $m_{j,k}(n)$, $k\in\ZZ$ belong to $\ZZ[{\bf w}]$.
It is easy to see that if $k$ is odd, or if $|k|>2j$, then $m_{j,k} \equiv 0$. It is also easy to see that 
\beq \label{obvious} m_{j,k} \=  A_{2j-1,k}\,. \eeq

\begin{lemma} The following identities hold true 
\begin{align}
& m_{j,\,-2}(n)  \= w_n \, w_{n-1} \, m_{j,2}(n-2)\,, \label{id00} \\
& m_{j,0}(n) \= m_{j-1,-2}(n) + m_{j-1,-2}(n+2) + (w_{n+1}+w_n) \, m_{j-1,0}(n)\,, \label{id01} \\
& m_{j,\,-2}(n)  - m_{j,\,-2}(n-2) - (w_{n-1}+w_{n-2}) \, \bigl(m_{j-1,-2}(n) -m_{j-1,-2}(n-2)\bigr)  \nn\\
& \qquad\qquad\qquad\qquad\qquad + w_{n-2} \, w_{n-3} \, m_{j-1,0}(n-4)- w_{n} \, w_{n-1} \, m_{j-1,0}(n) \=   0 \,. \label{id02}
\end{align}
\end{lemma}
\pf Comparing the constant terms of the identity 
\beq\label{LLLLL}
L^j \= L^{j-1} \, L \= L \, L^{j-1}
\eeq
we obtain that 
\begin{align}
m_{j,0}(n) & \= m_{j-1,-2}(n) +  (w_{n+1}+w_n) \, m_{j-1,0}(n) + w_{n+2} \, w_{n+1} \, m_{j-1,2}(n) \nn\\
& \= m_{j-1,-2}(n+2) + (w_{n+1}+w_n) \, m_{j-1,0}(n) + w_n \, w_{n-1} \, m_{j-1,2}(n-2)\,. \nn
\end{align}
This proves \eqref{id00}--\eqref{id01}. 
Similarly, comparing the coefficients of~$\Lambda^{-2}$ of~\eqref{LLLLL} we obtain  
\begin{align}
m_{j,-2}(n) & \= m_{j-1,-4}(n) +  (w_{n-1}+w_{n-2}) \, m_{j-1,-2}(n) + w_{n} \, w_{n-1} \, m_{j-1,0}(n) \nn\\
& \= m_{j-1,-4}(n+2) + (w_{n+1}+w_n) \, m_{j-1,-2}(n) + w_n \, w_{n-1} \, m_{j-1,0}(n-2) \,, \nn
\end{align}
which implies identity~\eqref{id02}. The lemma is proved.
\epf

\begin{lemma} The following identities hold true
\begin{align}
& A_{\ell,-1}(n) \= w_n \, A_{\ell,1}(n-1)\,, \label{id0p} \\
& A_{\ell,0}(n) \= w_{n+1} \, A_{\ell-1,1}(n) + w_n \, A_{\ell-1, 1}(n-1) \,,\label{id1p}\\
& w_n \, A_{\ell,1}(n-1) - w_{n+1} \, A_{\ell,1}(n) + w_{n+1}\, A_{\ell-1,0}(n-1) - w_n \, A_{\ell-1,0}(n-1) \= 0 \,, \label{id2p} \\
& A_{\ell,0}(n+1)  - A_{\ell,0}(n) \= w_{n+2}\, A_{\ell,2}(n) -  w_n\, A_{\ell,2}(n-1) \,.  \label{idpkey}
\end{align}
\end{lemma}
\pf Identities \eqref{id0p}--\eqref{id2p} are contained in the Lemma 2.2.1 of \cite{DuY1} (see the proof therein). 
Identity \eqref{idpkey} follows from comparing coefficients of $\Lambda$ on the both sides of the following identity:
$$
P^{\ell+1} P \= P P^{\ell+1}\,.
$$
The lemma is proved.
\epf 

Taking $\ell=2j-1$ in identity~\eqref{idpkey} and using~\eqref{obvious} we obtain 
\beq \label{keyid} m_{j,0}(n+1)  - m_{j,0}(n) \= w_{n+2}\, m_{j,2}(n) -  w_n\, m_{j,2}(n-1) \,. \eeq
We call this identity the {\it key identity}.  
It should be noted that the above identities \eqref{obvious}--\eqref{id02}, \eqref{id0p}--\eqref{idpkey} 
hold in $\ZZ[{\bf w}]$ absolutely (namely, the validity does not require that $w_n$ is a solution of the discrete KdV hierarchy), 
because they are nothing but properties of the operators $P$ and $L$. 

\subsection{Proof of Lemma~\ref{equivdkdv}} \label{section2point2}
Note that this lemma means the following: if $w_n=w_n({\bf s})$ satisfies~\eqref{defred}, 
then it satisfies~\eqref{dkdv}; {\it vice versa}.
Firstly, let $w_n=w_n({\bf s})$ be an arbitrary solution to~\eqref{defred}, 
i.e., $$\frac{\p P}{\p s_j} \=  \bigl[ A_{2j-1} \,,\, P \bigr]$$ for all $j\geq 1$. Since $L=P^2$ we have
$$
\frac{\p L}{\p s_j} \= P  \frac{\p P}{\p s_j}  +  \frac{\p P}{\p s_j} P \= P \, \bigl[ A_{2j-1} \,,\, P \bigr] + \bigl[ A_{2j-1} \,,\, P \bigr]  \, P \= [A_{2j-1}, \,L]\,.
$$
Secondly, let $w_n=w_n({\bf s})$ be an arbitrary solution to \eqref{dkdv}, namely, it satisfies that 
\begin{align}
& \frac{\p (w_{n+1} + w_n)}{\p s_j}   \=  w_{n+2} \, w_{n+1} \, m_{j,2}(n)  \, -  \, w_{n} \, w_{n-1}\, m_{j,2}(n-2)\,, \label{dkdv-e1}\\
& \frac{\p (w_n w_{n-1})}{\p s_j} \= w_{n} \, w_{n-1} \, \bigl( m_{j,0}(n) \,-\, m_{j,0}(n-2) \bigr)\,. \label{dkdv-e2}
\end{align}
Identity \eqref{dkdv-e1} implies that 
\begin{align}
(\Lambda+1)\frac{\p w_{n}}{\p s_j}  \= &  w_{n+2} \, w_{n+1} \, m_{j,2}(n)  -  w_{n+1} \, w_{n} \, m_{j,2}(n-1)  \nn\\ 
&  + w_{n+1} \, w_{n} \, m_{j,2}(n-1) \, -  \, w_{n} \, w_{n-1}\, m_{j,2}(n-2) \nn\\
 \= & w_{n+1} \bigl(m_{j,0}(n+1)  - m_{j,0}(n)\bigr) + w_n \bigl(m_{j,0}(n)  - m_{j,0}(n-1)\bigr) \,,   \nn
\end{align}
where we have used identity~\eqref{keyid}. Identity~\eqref{dkdv-e2} implies that 
$$
w_{n} \frac{\p w_{n+1}}{\p s_j} + w_{n+1} \frac{\p w_{n}}{\p s_j}   \= w_{n+1} \, w_n \, \bigl( m_{j,0}(n+1) - m_{j,0}(n) \bigr)+ w_{n+1} \, w_{n} \bigr(m_{j,0}(n) -  m_{j,0}(n-1) \bigr) \,.
$$
Combining the above two identities and assuming that $w_n\not\equiv w_{n+1}$ yields
\beq\label{discKdVflow}
\frac{\p w_n} {\p s_j} \= w_n \, \bigl( m_{j,0}(n)-m_{j,0}(n-1) \bigr) \= {\rm Coef}_{\Lambda^{-1}} \bigl[A_{2j-1},P\bigr]\,.
\eeq
(One can see from~\eqref{dkdv-e1} that solutions satisfying $w_n\equiv w_{n+1}$ 
are independent of~${\bf s}$. Therefore these trivial solutions also satisfy~\eqref{defred}.)
The proposition is proved.
\epf

\subsection{Lax pairs in matrix form} In this subsection we write the 
scalar Lax pairs \eqref{scalarl}--\eqref{scalara} into matrix form. 
The following lemma plays an important role. 
\begin{lemma} The wave function $\psi_n$ satisfies that
\beq\label{scalarpsij}
\frac{\p \psi_n}{\p s_j} \= \lambda^j \, \psi_n + \sum_{i=1}^j \lambda^{j-i} \, \bigl(m_{i-1,-2} \, \psi_n - w_n \, w_{n-1} \, m_{i-1,0} \, \psi_{n-2} \bigr) \, , \quad j\geq 1\,.
\eeq
\end{lemma}
\pf
We have for any $j\geq 1$
\begin{align}
\bigl(L^j\bigr)_+  & \= \bigl(L^{j-1} L\bigr)_+ \= \bigl(L^{j-1}\bigr)_+ \, L_+ \+ \Bigl(\bigl(L^{j-1}\bigr)_- \, L\Bigr)_+ \+ \Bigl( \bigl(L^{j-1}\bigr)_+ \, L_-\Bigr)_+\nn\\
& \= \bigl (L^{j-1}\bigr)_+  \, L \,-\, \Bigl(\bigl(L^{j-1}\bigr)_+ \, L_-\Bigr)_- \+ \Bigl( \bigl(L^{j-1}\bigr)_- \, L\Bigr)_+ \nn\\
& \= \bigl (L^{j-1}\bigr)_+  \, L \+ m_{j-1,-2}  \,-\, w_n \, w_{n-1} \, m_{j-1,0} \, \Lambda^{-2}\,.  \nn
\end{align}
In the above derivations it is understood that $L = L(n)$ and $m_{j,k} = m_{j,k}(n)$. Therefore,
$$
A_{2j-1} \= \bigl(L^j\bigr)_+ \= L^j \+ \sum_{i=1}^j \bigl(m_{i-1,-2} - w_n \, w_{n-1} \, m_{i-1,0} \, \Lambda^{-2} \bigr) \, L^{j-i}\,, \qquad \forall \, j\geq 0.
$$
The lemma is proved.
\epf

\begin{lemma}\label{lemmaPsis} The vector-valued wave function 
$\Psi_n=\bigl(\psi_n,\psi_{n-2}\bigr)^T$ satisfies that 
\beq\label{matrixaj}
\frac{\p \Psi_n}{\p s_j} \= V_j(n) \, \Psi_n \,, \qquad j\geq 1\,, 
\eeq
where $V_j(n)$ are the following $2\times2$ matrices
\beq\label{defvjn}
V_j(n) \:= \Biggl(\begin{array} {cc} \lambda^j + \sum_{i=1}^j \lambda^{j-i} \, m_{i-1,-2}(n)  & -w_n \, w_{n-1}\sum_{i=1}^j \lambda^{j-i}   \, m_{i-1,0}(n) \\ 
\sum_{i=1}^j \lambda^{j-i} m_{i-1,0}(n-2) &   m_{j,0}(n-2)-\sum_{i=1}^j \lambda^{j-i} m_{i-1,-2}(n) \end{array}\Biggr) \,.
\eeq
\end{lemma}
\pf Equation~\eqref{matrixaj} follows straightforwardly from~\eqref{scalarpsij} and~\eqref{scalarl}.
\epf

We therefore arrive at

\begin{prop} \label{matrixlax}
The discrete KdV hierarchy are the compatibility conditions of~\eqref{matrixl} and~\eqref{matrixaj}:
$$
\frac{\p U_n}{\p s_j} \= V_j(n+2) \, U_n \,-\, U_n \, V_j(n) \,, \qquad j=1,2,3,\cdots.
$$
\end{prop}

\section{Tau-structure for the discrete KdV hierarchy} \label{MRsection}
In this section, we use the MR method to study the tau-structure of the discrete KdV hierarchy; 
in particular, we will prove Proposition~\ref{main1}.
The notations about the matrix-resolvents are the same as in the Introduction.
\subsection{The MR recursive relations}
Write 
\beq
\alpha_n \= \sum_{j\geq 0} \frac{a_{n,j}}{\lambda^{j+1}},\qquad \gamma_n \= \sum_{j\geq 0} \frac{c_{n,j}}{\lambda^{j+1}} \,.
\eeq
Then we find that $a_{n,j},\,c_{n,j}$ satisfy 
\begin{align}
& c_{n,j+1} \= (w_{n-1}+w_{n-2}) \, c_{n,j} \+ a_{n,j} + a_{n-2,j}\,, \label{mr1}\\
& a_{n,j+1} \,-\, a_{n+1,j+1} \+ (w_{n+1}+w_n) (a_{n+2}-a_{n,j}) \+ w_{n+1} \,w_n \, c_{n+4,j} \,-\, w_n \, w_{n-1} \, c_{n,j} \= 0\,, \label{mr2}\\
& a_{n,j} \= \sum_{i=0}^{j-1} \bigl(w_n \, w_{n-1} \, c_{n,i} \, c_{n,j-1-i} - a_{n,i} \, a_{n,j-1-i}\bigr)\label{mr3}
\end{align}
as well as
\beq
a_{n,0} \= 0 \,,\qquad c_{n,0} \= 1 \,.
\eeq

\begin{lemma}
The basic resolvent of ${\mathcal L}$ exists and is unique. 
\end{lemma}
\pf 
Observe that multiplying~\eqref{rrr2} and~\eqref{rrr3} gives~\eqref{rrr4}.
This proves existence of $R_n$. 
Uniqueness follows directly from the MR recursive relations \eqref{mr1}--\eqref{mr3}, as we can solve $a_{n,j},\, c_{n,j}$ 
uniquely in an algebraic way for all $j\geq 1$. The lemma is proved. 
\epf

For the reader's convenience we give in below the first few terms of the basic resolvent of~$\mathcal{L}$:
$$
R_n(\lambda) \= \begin{pmatrix}
1+ \frac{w_{n-1} w_n}{\lambda^2}+\cdots 
& -\frac{w_{n-1} w_n}{\lambda}-\frac{w_{n-1} (w_n+w_{n+1}) w_n}{\lambda ^2}+ \cdots \\ \\
 \frac{1}{\lambda} + \frac{w_{n-2}+w_{n-1}}{\lambda ^2} + \cdots & -\frac{w_{n-1} w_n}{\lambda^2}+ \cdots \\
\end{pmatrix} \,.
$$

\subsection{MR and the discrete KdV flows} 
In this subsection we use the basic MR to express the discrete KdV flows. (We would like to mention that the materials that 
we give in this subsection are rather standard.)
Let $R_n$ be the basic matrix resolvent of~$\mathcal{L}$. 
\begin{lemma} \label{lemmaca}
The following formulae hold true:
\begin{align}
& c_{n,j} \= m_{j,0}(n-2) \,, \label{cm} \\
& a_{n,j} \= m_{j,-2}(n) \,. \label{am} 
\end{align}
\end{lemma}
\pf
By identifying their recursive relations as well as the initial values of the recursions.
\epf

It follows from the above Lemma~\ref{lemmaca} that the matrices $V_j(n)$ defined in~\eqref{defvjn} have the following expressions:
\beq\label{vjn}
V_j(n) \= \bigl(\lambda^j R_n \bigr)_+ \+  \biggl(\begin{array} {cc} 0 & 0 \\ 0 &  c_{n,j}\\  \end{array}\biggr) \,, 
\eeq
where $``+"$ means taking the polynomial part in~$\lambda$ (including the constant term). 

\subsection{Loop operator}
Introduce a linear operator $\nabla(\lambda)$ by
\beq\label{loop}
\nabla(\lambda) \:= \sum_{j\geq 1} \frac1{\lambda^{j+1}} \frac{\p }{\p s_j} \,.
\eeq
It readily follows from equation \eqref{vjn} that 
$$
\nabla(\mu) \, \Psi_n(\lambda) \= \biggl[\frac{R_n(\mu) }{\mu-\lambda}+ Q_n(\mu)\biggr] \, \Psi_n(\lambda) \,, 
$$
where 
$$
Q_n(\mu) \:= -\frac{I}{\mu} + \begin{pmatrix}  0 & 0 \\ 0 &  \gamma_n(\mu)\\  \end{pmatrix} \,.
$$
\begin{lemma} The following formula holds true:
\beq
\nabla(\mu) \, R_n(\lambda) \= \frac1{\mu-\lambda} \bigl[R_n(\mu), R_n(\lambda)\bigr] \+ \bigl[Q_n(\mu), R_n(\lambda)\bigr] \,.
\eeq
\end{lemma}

\subsection{From MR to tau-function}
The MR allows us to define tau-function of an arbitrary solution of the discrete KdV hierarchy.
Recall that a family of elements $\Omega_{p;q}(n) \in \ZZ[{\bf w}]$, $p,q\geq 1$ are called a tau-structure of the discrete KdV hierarchy if 
\begin{align}
& \Omega_{p;q}(n) \= \Omega_{q;p}(n) \, , \qquad \forall\, p,q\geq 1
\end{align}
and for an arbitrary solution $w_n=w_n({\bf s})$ of the discrete KdV hierarchy
\begin{align}
& \frac{\p \Omega_{p;q}(n)}{\p s_r} \= \frac{\p \Omega_{p;r}(n)}{\p s_q}\,,\qquad \forall \, p,q,r\geq 1 \, .
\end{align}

\begin{defi} 
Define $\Omega_{i;j}(n)$, $i,j\geq 1$ via the generating series 
\beq\label{defOmega}
\sum_{i,j\geq 1} \Omega_{i;j}(n) \, \lambda^{-i-1} \mu^{-j-1} \= \frac{\tr \, \bigl(R_n(\lambda) R_n(\mu)\bigr) -1 }{(\lambda-\mu)^2 } \,.
\eeq
\end{defi}
\begin{lemma}\label{taustrulemma}
The $\Omega_{i;j}(n)$, $i,j\geq 1$ \eqref{defOmega} are well-defined, and live in $\ZZ[{\bf w}]$. Moreover,  they form a tau-structure of the discrete KdV hierarchy.
\end{lemma}
\pf 
The proof is almost identical with the one for the Toda lattice hierarchy~\cite{DuY1} (or the one for the Drinfeld--Sokolov 
hierarchies~\cite{BDY2}); details are omitted here.
\epf

\pf of Lemma~\ref{fromMRtotau}. \quad By Lemma~\ref{taustrulemma}, it suffices to prove the compatibility between \eqref{d1}--\eqref{d3}. 

Firstly, on one hand,
\begin{align}
& \sum_{i,j\geq 1} \lambda^{-i-1} \mu^{-j-1} \Bigl[\Omega_{i;j}(n+2)-\Omega_{i;j}(n)\Bigr] \nn\\
& \= \frac{\tr \, \bigl(R_{n+2}(\lambda) R_{n+2}(\mu)\bigr) - \tr \, \bigl(R_n(\lambda) R_n(\mu)\bigr) }{(\lambda-\mu)^2 } \nn\\
& \= \frac{ (1+2\alpha_n(\lambda))\, \gamma_{n+2}(\mu) - (1+2\alpha_n(\mu)) \, \gamma_{n+2}(\lambda) }{\lambda-\mu} - \gamma_{n+2}(\lambda)\gamma_{n+2}(\mu)\,. \nn
\end{align}
On the other hand,
\begin{align}
\nabla(\mu) \bigl[R_{n+2}(\lambda)\bigr]_{21} \=  \frac{\bigl(1+2\alpha_{n+2}(\mu)\bigr)\gamma_{n+2}(\lambda)- \bigl(1+2\alpha_{n+2}(\lambda)\bigr) \, \gamma_{n+2}(\mu)}{\lambda-\mu} + \gamma_{n+2}(\lambda)\gamma_{n+2}(\mu)\,. \nn
\end{align}
Hence by using \eqref{rrr2} we find that 
\beq\label{already}
\sum_{i,j\geq 1} \lambda^{-i-1} \mu^{-j-1} \Bigl[\Omega_{i;j}(n+2)-\Omega_{i;j}(n)\Bigr] \= \nabla(\mu) \bigl[R_{n+2}(\lambda)\bigr]_{21} \,. 
\eeq
This proves the compatibility between \eqref{d1} and \eqref{d2}. 

Secondly, on one hand, 
\begin{align}
& \sum_{i,j\geq 1} \lambda^{-i-1} \mu^{-j-1} \Bigl[\Omega_{i;j}(n+2)+ \Omega_{i;j}(n-1)-\Omega_{i;j}(n+1)- \Omega_{i;j}(n)\Bigr]  \nn\\
&\= \sum_{i,j\geq 1} \lambda^{-i-1} \mu^{-j-1} \Bigl[\Omega_{i;j}(n+2)- \Omega_{i;j}(n)\Bigr] - \sum_{i,j\geq 1} \lambda^{-i-1} \mu^{-j-1} \Bigl[\Omega_{i;j}(n+1)- \Omega_{i;j}(n-1)\Bigr] \,.\nn
\end{align}
On the other hand,
\begin{align}
\nabla (\mu) \nabla(\lambda) \log w_n \= \nabla(\mu) \Bigl[ \gamma_{n+2}(\lambda)-\gamma_{n+1}(\lambda) \Bigr] \= \nabla(\mu) \gamma_{n+2}(\lambda)- \nabla(\mu) \gamma_{n+1}(\lambda)\,.
\end{align}
Using \eqref{already} we find 
\begin{align}
& \sum_{i,j\geq 1} \lambda^{-i-1} \mu^{-j-1} \Bigl[\Omega_{i;j}(n+2)+ \Omega_{i;j}(n-1)-\Omega_{i;j}(n+1)- \Omega_{i;j}(n)\Bigr]  \= \nabla (\mu) \nabla(\lambda) \log w_n \,.
\end{align}
This proves compatibility between \eqref{d1} and \eqref{d3}. Thirdly, the following identity
$$
\nabla(\lambda) \log w_n \=  \gamma_{n+2}(\lambda)-\gamma_{n+1}(\lambda)
$$
shows the compatibility between \eqref{d2} and \eqref{d3}. The proposition is proved.
\epf

\subsection{Generating series of multi-point correlations functions}\label{Generatingsubsection}
For an arbitrary solution $w_n=w_n({\bf s})$ to the discrete KdV hierarchy, let $\tau_n^{\textsc{\tiny\rm dKdV}}=\tau_n^{\textsc{\tiny\rm dKdV}}({\bf s})$ 
denote the tau-function of this solution. The logarithmic derivatives 
$$\frac{\p^k \log \tau_n^{\textsc{\tiny\rm dKdV}}({\bf s})}{\p s_{j_1}\dots \p s_{j_k}}\,, \quad j_1,\dots,j_k\geq 1, ~ k\geq 1$$
can be called the $k$-point correlation 
functions\footnote{We can say in a more accurate sense that the logarithmic derivatives are 
identified with the correlation functions, where the latter are defined as abstract differential polynomials; 
see for example~\cite{DYZ2} for the details.} of the solution $w_n=w_n({\bf s})$.

\pf of Proposition~\ref{main1}. \quad 
The proof can be achieved by the mathematical induction, as in \cite{BDY1}; we hence omit the details. 
\epf

We see from Proposition~\ref{main1} that the logarithmic derivatives 
$\frac{\p^k \log \tau_n^{\textsc{\tiny\rm dKdV}}({\bf s})}{\p s_{j_1}\dots \p s_{j_k}}$ with $k\geq 2$ 
all live in $\ZZ[{\bf w}]$, as their generating series are expressed by MR via algebraic 
manipulations; this simple fact agrees with footnote~2 (and can be of course deduced from other techniques).

\section{Proof of Theorem~\ref{maintheorem}} \label{SectionMaintheorem}
The goal of this section is to prove Theorem~\ref{maintheorem}.
\subsection{Review of the MR approach to the Toda lattice hierarchy}
Denote 
$$
\mathcal{P} \:= \Lambda + v_n^{\textsc{\tiny\rm Toda}} + w_n^{\textsc{\tiny\rm Toda}} \Lambda^{-1}\,,\qquad  \quad
\mathcal A_\ell \:= \bigl(\mathcal P^{\ell+1}\bigr)_+\,, \quad \ell\geq 0. 
$$
The Toda lattice hierarchy is defined as the following system of commuting flows
\beq
\frac{\p \mathcal P}{\p t_\ell} \= \Bigl[ \mathcal A_\ell\,, \, \mathcal P \Bigr]\,, \qquad \ell\geq 0\,.
\eeq

Let us briefly review part of the results of~\cite{DuY1}. 
Introduce 
$\mathcal U_n \= \biggl(\begin{array} {cc} v_n^{\textsc{\tiny\rm Toda}} -\lambda & w_n^{\textsc{\tiny\rm Toda}} \\ -1 & 0\\  \end{array}\biggr)$. 
The basic resolvent  ${\mathcal R}_n$ associated to
$\mathcal  P^{\rm M} := \Lambda + \mathcal U_n$ is defined as the unique 
solution in 
${\rm Mat}\bigl(2,\ZZ\bigl[{\bf v}^{\textsc{\tiny\rm Toda}},{\bf w}^{\textsc{\tiny\rm Toda}}\bigr][[\lambda^{-1}]]\bigr)$ 
to the problem:
\begin{align}
& \mathcal R_{n+1} \, \mathcal U_n - \mathcal U_n \, \mathcal R_n \= 0\,,\\
& \mathcal R_n \= \begin{pmatrix}  1  & 0 \\ 0 & 0 \\  \end{pmatrix}  + \mathcal{O}\bigl(\lambda^{-1}\bigr) \,,  \\ 
& \tr \, \mathcal R_n \= 1\,, \qquad {\rm det} \, \mathcal R_ n \= 0\,. 
\end{align} 
Write 
\beq
\mathcal R_n (\lambda) \=  \begin{pmatrix}  1+ \mathcal{A}_n(\lambda)  & \mathcal{B}_n(\lambda) \\ \mathcal{G}_n(\lambda) & - \mathcal{A}_n(\lambda) \\  \end{pmatrix} \,, \qquad
\mathcal{A}_n,\mathcal{B}_n,\mathcal{G}_n \in \ZZ[{\bf v}^{\textsc{\tiny\rm Toda}},{\bf w}^{\textsc{\tiny\rm Toda}}][[\lambda^{-1}]]\,.
\eeq
Then $\mathcal{A}_n,\mathcal{B}_n,\mathcal{G}_n$ satisfy that 
\begin{align}
& \mathcal{B}_n \= -w_n^{\textsc{\tiny\rm Toda}} \,\mathcal{G}_{n+1}  \label{rr1} \\
& \mathcal{A}_{n+1} +\mathcal{A}_n+1 \= \mathcal{G}_{n+1}  \, (\lambda-v_n^{\textsc{\tiny\rm Toda}}) \label{rr2} \\
& (\lambda-v_n^{\textsc{\tiny\rm Toda}}) (\mathcal{A}_n-\mathcal{A}_{n+1}) \= w_n^{\textsc{\tiny\rm Toda}} \, \mathcal{G}_n - w_{n+1}^{\textsc{\tiny\rm Toda}} \, \mathcal{G}_{n+2}\label{rr3} \\
& \mathcal{A}_n + \mathcal{A}_n^2 \= \mathcal{B}_n \, \mathcal{G}_n\,. \label{rr4}
\end{align}

The following lemma was proven in \cite{DuY1}. 
\begin{lemma}[\cite{DuY1}] \label{lemmatwo} For an arbitrary solution $v_n^{\textsc{\tiny\rm Toda}}=v_n^{\textsc{\tiny\rm Toda}}({\bf t})$, $w_n^{\textsc{\tiny\rm Toda}}=w_n^{\textsc{\tiny\rm Toda}}({\bf t})$ to the Toda lattice hierarchy there exists a function $\tau_n^{\textsc{\tiny\rm Toda}}({\bf t})$ such that
\begin{align}
& \sum_{i,\, j\geq 0} \frac{1}{\lambda^{i+2} \mu^{j+2}} \frac{\partial^2\log\tau^{\textsc{\tiny\rm Toda}}_n({\bf t})}{\partial t_i \,\partial t_j} \= \frac{{\rm tr}\, \R_n({\bf t},\lambda) \R_n({\bf t},\mu)-1}{(\lambda-\mu)^2}
\label{taun1} \\
& \frac1\lambda + \sum\limits_{i\geq 0} \frac1{\lambda^{i+2}}\frac{\pal}{\pal t_i} \log \frac{\tau^{\textsc{\tiny\rm Toda}}_{n+1}({\bf t})}{\tau^{\textsc{\tiny\rm Toda}}_n({\bf t})}  \=  \left[ \R_{n+1}({\bf t},\lambda)\right]_{21}
\label{taun2} \\
& \frac{\tau^{\textsc{\tiny\rm Toda}}_{n+1}({\bf t}) \tau^{\textsc{\tiny\rm Toda}}_{n-1}({\bf t})}{\tau^{\textsc{\tiny\rm Toda}}_n({\bf t})^2} \= w_n. \label{taun3}
\end{align}
The function $\tau_n^{\textsc{\tiny\rm Toda}}({\bf t})$ is uniquely determined 
by the solution $v_n^{\textsc{\tiny\rm Toda}}({\bf t})$, $w_n^{\textsc{\tiny\rm Toda}}({\bf t})$ up to 
$$
\tau_n^{\textsc{\tiny\rm Toda}}({\bf t}) \mapsto e^{a_0 + a_1 n + \sum_{j\geq 0} b_j t_j} \tau^{\textsc{\tiny\rm Toda}}_n({\bf t})
$$
for some constants $a_0$, $a_1$, $b_0$, $b_1$, $b_2$, $\dots$..
\end{lemma}

In \cite{DuY1} the $\tau_n^{\textsc{\tiny\rm Toda}}({\bf t})$ is called the {\it tau-function} of the solution $v_n^{\textsc{\tiny\rm Toda}}({\bf t})$, $w_n^{\textsc{\tiny\rm Toda}}({\bf t})$ to the Toda lattice hierarchy. The logarithmic derivatives of $\tau_n^{\textsc{\tiny\rm Toda}}({\bf t})$
$$
\frac{\p^k\log \tau_n^{\textsc{\tiny\rm Toda}}({\bf t})}{\p t_{i_1}\dots \p t_{i_k}}\,, \qquad i_1,\dots,i_k\geq 0\,, k\geq 1
$$
can be called $k$-point correlations functions (cf. footnote~2) of the Toda lattice hierarchy. Define
$$
C_k( \lambda_1,\dots,\lambda_k; n; {\bf t}) \:=  
\sum_{i_1,\dots,i_k\geq 0} \frac1{\lambda^{i_1+2}_1\cdots \lambda^{i_k+2}_{k}} 
\frac{\p^k\log \tau_n^{\textsc{\tiny\rm Toda}}({\bf t})}{\p t_{i_1}\dots \p t_{i_k}}\,.
$$

\subsection{Reduction to the discrete KdV hierarchy} \label{reductionsection}
Now consider solutions to the Toda lattice hierarchy in the ring $\CC[[t_0,t_1,\dots]]\otimes \V$, 
where $\V$ is any ring of functions of~$n$, closed under~$\Lambda$ and~$\Lambda^{-1}$.
These solutions can be specified by (i.e. are in 1-1 correspondence to) the initial value:
$$
f(n)= v_n^{\textsc{\tiny\rm Toda}}({\bf t}={\bf 0})\,,\quad g(n)= w_n^{\textsc{\tiny\rm Toda}}({\bf t}={\bf 0})\,.
$$

Let us explain how a subset of solutions to the Toda lattice hierarchy be reduced to solutions of the discrete KdV hierarchy. 
On one hand, let $v_n^{\textsc{\tiny\rm Toda}}=v_n^{\textsc{\tiny\rm Toda}}({\bf t})$, 
$w_n^{\textsc{\tiny\rm Toda}}=w_n^{\textsc{\tiny\rm Toda}}({\bf t})$ be an arbitrary solution in $\CC[[t_0,t_1,\dots]]\otimes \V$ of the Toda lattice hierarchy satisfying the following type of initial conditions 
$$
f(n) \, \equiv \, 0 \,.
$$
It follows that 
 \begin{align}
 v_n^{\textsc{\tiny\rm Toda}}\big|_{t_0=t_2=t_4=\cdots=0} \, \equiv \, 0\,, \qquad( \forall\,  n,t_1,t_3,t_5,\cdots). \label{reduction1}
 \end{align}
This further implies that 
the commuting flows $\frac{\p w_n^{\textsc{\tiny\rm Toda}}(\bt)}{\p t_{2j-1}}\big|_{t_0=t_2=t_4=\dots=0}\, (j\geq 1)$ are decoupled, 
namely, there are no $v_n^{\textsc{\tiny\rm Toda}}$-dependence in these flows (of course when restricting to $t_0=t_2=t_4=\dots=0$). 
Moreover, these flows   
 coincide with the discrete KdV hierarchy \eqref{defred}. Therefore if we define 
 \beq\label{defwnfromreduction}
 w_n({\bf s}) \:= w_n^{\textsc{\tiny\rm Toda}}({\bf t})\big|_{t_{2i-1}=s_i, \, t_{2i-2}=0, \,i\geq 1} \,, 
 \eeq
 then $w_n=w_n({\bf s})$ is a solution to the discrete KdV hierarchy.
On the other hand, let $w_n=w_n({\bf s})$ be an arbitrary solution to the discrete KdV hierarchy in the ring $\CC[s_1,s_2,\dots]\otimes \V$. Let $g(n)$ denote its initial value, i.e. 
$g(n):=w_n({\bf s}={\bf 0})$. Define 
$v_n^{\textsc{\tiny\rm Toda}}(\bt)\,, w_n^{\textsc{\tiny\rm Toda}}(\bt)$ as the unique solution in $\CC[[t_0,t_1,\dots]]\otimes \V$ to the Toda 
lattice hierarchy with $(f(n)\equiv 0, g(n))$ as the initial value. Then $w_n^{\textsc{\tiny\rm Toda}}({\bf t})|_{t_{2i-1}=s_i, \, t_{2i-2}=0, \,i\geq 1} = w_n({\bf s})$. 

Hence the correspondence between solutions of the discrete KdV hierarchy 
and a suitable subset of solutions of the Toda lattice hierarchy has been established. 

For a solution $\bigl(v_n^{\textsc{\tiny\rm Toda}}(\bt), w_n^{\textsc{\tiny\rm Toda}}(\bt)\bigr)$ to the 
Toda lattice hierarchy satisfying $v_n^{\textsc{\tiny\rm Toda}}({\bf 0})\equiv 0~(\forall\,n)$, let $\tau_n^{\textsc{\tiny\rm Toda}}(\bt)$ denote the tau-function 
of this solution. Define $w_n({\bf s})$ as in \eqref{defwnfromreduction}, and
\begin{align}
& \tau_n ({\bf s}) \:= \tau_n^{\textsc{\tiny\rm Toda}}(t_0 = 0, t_1 = s_1, t_2 =0, t_3 =s_2, \cdots ) \,. \nn
\end{align}
Then we know that the function $w_n=w_n({\bf s})$ satisfies the discrete KdV hierarchy \eqref{defred}, and that
\beq\label{wtaudy}
w_n({\bf s}) \= \frac{\tau_{n+1}({\bf s}) \, \tau_{n-1} ({\bf s})}{\tau_n^2({\bf s})}  \,. 
\eeq
As indicated above, all solutions of the discrete KdV hierarchy can be obtained from this way. 
\begin{defi}
We call $\tau_n({\bf s})$ the tau-function reduced from the 
Toda lattice hierarchy of the solution $w_n=w_n({\bf s})$ 
to the discrete KdV hierarchy.
\end{defi}

Introduce the notations: 
\begin{align}
& A_n(\lambda) \:= \mathcal{A}_n(\lambda)|_{v_n^{\textsc{\tiny\rm Toda}} \equiv 0, \, w_n^{\textsc{\tiny\rm Toda}} \equiv w_n}\,, \nn\\
& B_n(\lambda) \:= \mathcal{B}_n(\lambda)|_{v_n^{\textsc{\tiny\rm Toda}} \equiv 0, \, w_n^{\textsc{\tiny\rm Toda}} \equiv w_n}\,, \nn\\
& G_n(\lambda) \:= \mathcal{G}_n(\lambda)|_{v_n^{\textsc{\tiny\rm Toda}} \equiv 0, \, w_n^{\textsc{\tiny\rm Toda}} \equiv w_n}\,. \nn
\end{align}
Clearly, $A_n,~B_n, ~G_n$ belong to $\ZZ[{\bf w}][[\lambda^{-1}]]$. Note that definitions of $A_n(\lambda),~B_n(\lambda),~G_n(\lambda)$ are in the absolute sense, 
namely, they do not depend on whether $w_n$ is a solution or not. 
\begin{lemma} \label{AREC}
The $A_n(\lambda)$ satisfies 
\beq\label{recursionA}
w_{n+1} \bigl( A_{n+2} (\lambda) + A_{n+1}(\lambda) +1 \bigr) - w_n \bigl( A_n(\lambda) + A_{n-1}(\lambda) + 1 \bigr) \= \lambda^2 \, \bigl( A_{n+1}(\lambda) - A_n(\lambda) \bigr)\,.
\eeq
\end{lemma}
\pf 
Following from \eqref{rr2} and \eqref{rr3} with $v_n^{\textsc{\tiny\rm Toda}}\equiv 0$. 
\epf

\subsection{Proof of Theorem~\ref{maintheorem}} 
Firstly, 
on one hand, it follows from the Lemma 1.2.3 of~\cite{DuY1} that 
\beq\label{mh75}
m_{j,0}(n;{\bf s}) \= \frac{\p }{\p s_{j}} \log \frac{ \tau_{n+1}({\bf s})}{\tau_n({\bf s})} \,, \qquad j\geq 1\,.
\eeq
On the other hand, from \eqref{d2} and \eqref{cm} we find
\beq
m_{j,0}(n;{\bf s}) \= \frac{\p }{\p s_{j}} \log \frac{ \tau_{n+2}^{\textsc{\tiny\rm dKdV}}({\bf s})}{\tau_n^{\textsc{\tiny\rm dKdV}}({\bf s})} \,, \qquad j\geq 1\,.
\eeq
Comparing the above two expressions we find 
\beq\label{ttd1}
\log \frac{ \tau_{n+1}({\bf s})}{\tau_n({\bf s})} - \log \frac{ \tau_{n+2}^{\textsc{\tiny\rm dKdV}}({\bf s})}{\tau_n^{\textsc{\tiny\rm dKdV}}({\bf s})} \= S(n) \,, 
\eeq
where $S(n)$ is some function depending only on~$n$.  Equation~\eqref{ttd1} implies that 
\beq\label{ttd12}
\log \tau_n({\bf s}) \,-\, (\Lambda +1) \, \log \tau_n^{\textsc{\tiny\rm dKdV}}({\bf s}) \= \widetilde S(n) \+ f({\bf s}) \,,
\eeq
where $\widetilde S(n)$ is some function depending only on~$n$, 
and $f({\bf s})$ is some function depending only on~${\bf s}$.

Secondly, it follows from \eqref{d3} and \eqref{wtaudy} that
\beq\label{ttd2}
\frac{\tau_{n+1}({\bf s}) \, \tau_{n-1} ({\bf s})}{\tau_n^2({\bf s})}  \= 
\frac{\tau^{\textsc{\tiny\rm dKdV}}_{n+2}({\bf s}) \, \tau^{\textsc{\tiny\rm dKdV}}_{n-1}({\bf s})}{ \tau^{\textsc{\tiny\rm dKdV}}_{n+1}({\bf s}) \, \tau^{\textsc{\tiny\rm dKdV}}_n({\bf s})}  \,.
\eeq 
Substituting \eqref{ttd12} in \eqref{ttd2} we find that $\widetilde S(n)$ can only be an affine function of~$n$, namely, 
\beq\label{diff}
\log \tau_n({\bf s}) \,-\, (\Lambda +1) \, \log \tau_n^{\textsc{\tiny\rm dKdV}}({\bf s}) \= \alpha \,n \+ \alpha' \+ f({\bf s})  \,, 
\eeq
where $\alpha$, $\alpha'$ are some constants independent of~$n$, ${\bf s}$. 

Thirdly, on one hand, using~\eqref{d1} we find 
\begin{align}
& \sum_{i,j\geq 1} \frac{\p^2 \log \tau_n^{\textsc{\tiny\rm dKdV}} ({\bf s})}{\p s_i \p s_j} \frac1{\lambda^{i+1} \mu^{j+1}} \nn\\
&   \qquad \quad \= \frac{\alpha_n(\lambda)+\alpha_n(\mu)+2\alpha_n(\lambda)\alpha_n(\mu) 
- w_n\, w_{n-1}\, \bigl(\gamma_{n}(\lambda)\gamma_{n+2}(\mu) +\gamma_n(\mu)\gamma_{n+2}(\lambda)\bigr)}{(\lambda-\mu)^2} \,. \nn
\end{align}
Therefore,
\begin{align}
& \sum_{i,j\geq 1}   \frac{\p^2 \log \tau_n^{\textsc{\tiny\rm dKdV}} ({\bf s})}{\p s_i \p s_j} \frac1{\lambda^{2i+1} \mu^{2j+1}} \nn\\
 & \=  \lambda\mu \frac{\alpha_n(\lambda^2)+\alpha_n(\mu^2)+2\alpha_n(\lambda^2)\alpha_n(\mu^2) - w_n\, w_{n-1}\, \bigl(\gamma_{n}(\lambda^2)\gamma_{n+2}(\mu^2) +\gamma_n(\mu^2)\gamma_{n+2}(\lambda^2)\bigr)}{(\lambda^2-\mu^2)^2} \nn\\
 & \;=:\;  W_2(\lambda,\mu;n,{\bf s}) \,. \nn
\end{align}
So
\[
 \sum_{i,j\geq 1} \biggl(\frac{\p^2 \log \tau_n^{\textsc{\tiny\rm dKdV}} ({\bf s})}{\p s_i \p s_j} 
+ \frac{\p^2 \log \tau_{n+1}^{\textsc{\tiny\rm dKdV}} ({\bf s})}{\p s_i \p s_j}  \biggr)\frac1{\lambda^{2i+1} \mu^{2j+1}} \= W_2(\lambda,\mu;n,{\bf s}) + W_2(\lambda,\mu; n+1,{\bf s}) \,. \]
On the other hand, for $P(n)=\Lambda+w_n\Lambda^{-1}$, recall the notation   
$$
P(n)^{\ell+1} \=  \sum_{k\in\mathbb{Z}} A_{\ell,k}(n) \, \Lambda^{k} \,, \qquad  \ell =-1,0,1,2,\cdots. 
$$
Using Lemma~\ref{lemmatwo} 
we have 
\begin{align}
C_2(\lambda, \mu; n; {\bf t}) 
&  \= \frac{A_n(\lambda) + A_n(\mu)+2A_n(\lambda)A_n(\mu)-w_n\bigl(G_{n+1}(\lambda)G_n(\mu) + G_{n+1}(\mu)G_n (\lambda)\bigr)}{(\lambda-\mu)^2} \,,  \nn
\end{align}
where $$A_n(\lambda) \= \sum_{\ell \geq 0} \frac{A_{\ell-1,-1}(n)}{\lambda^{\ell+1}} \,, \qquad G_n(\lambda) \= \sum_{\ell \geq 0} \frac{A_{\ell-1,0}(n-1)}{\lambda^{\ell+1}} \,.$$
 Taking $$t_{2i-2} = 0, ~ t_{2i-1} = s_i ~ (i\geq 1)$$ we have
\beq\label{GP1}
G_n(\lambda) \= \sum_{j\geq 0}   \frac{A_{2j-1,0}(n-1)}{\lambda^{2j+1}} \= \sum_{j\geq 0}   \frac{m_{j,0}(n-1)}{\lambda^{2j+1}} 
\= \sum_{j\geq 0}   \frac{c_{n+1,j}}{\lambda^{2j+1}} \= \lambda\,\gamma_{n+1}(\lambda^2) \,. 
\eeq
It follows from~\eqref{rr2}, \eqref{GP1}, and respectively~\eqref{rrr2}, that 
\begin{align}
& A_n(\lambda)  \= (\Lambda+1)^{-1} \Bigl( \lambda^2 \gamma_{n+2} (\lambda^2) - 1 \Bigr) \= \lambda^2 \, (\Lambda+1)^{-1} \, \gamma_{n+2} (\lambda^2) - \frac12  \,, \label{Aga}\\
& \alpha_n(\lambda) 
\= (\Lambda^2+1)^{-1} \Bigl( (\lambda-w_{n+1}-w_n) \, \gamma_{n+2}(\lambda)\Bigr)- \frac12 \,. \label{aga}
\end{align}
\begin{lemma} \label{usefullemma} 
The following identities hold true:
\begin{align}
& \gamma_n(\lambda^2) \= \frac{ A_{n-1}(\lambda) + A_{n-2}(\lambda) + 1}{\lambda^2} \,, \label{g1}\\
& G_n(\lambda) \= \frac{ A_{n}(\lambda) + A_{n-1}(\lambda) + 1}{\lambda}  \,, \label{G1}\\
& \alpha_n(\lambda^2) \=  A_{n-1}(\lambda) - \frac{w_{n-1}}{\lambda^2} \, \bigl(A_{n-1}(\lambda) + A_{n-2}(\lambda) + 1\bigr)   \,. \label{a1}
\end{align}
\end{lemma}
\pf
Identities \eqref{g1}, \eqref{G1} are easy consequences of \eqref{Aga}, \eqref{GP1}. 

Note that identity~\eqref{keyid} implies that 
\begin{align}
\alpha_n(\lambda^2) 
& \= \frac12 \Bigl( w_{n-1} \, \gamma_{n+1}(\lambda^2) - w_{n-2} \, \gamma_{n-1}(\lambda^2)  + (\lambda^2-2w_{n-1})  \, \gamma_n(\lambda^2) -1 \Bigr) \nn\\
& \= \frac12 \Bigl( w_{n-1} \, \frac{ A_{n}(\lambda) + A_{n-1}(\lambda) + 1}{\lambda^2}- w_{n-2} \, \frac{ A_{n-2}(\lambda) + A_{n-3}(\lambda) + 1}{\lambda^2}  \nn\\
&\qquad\qquad + (\lambda^2-2w_{n-1})  \, \frac{ A_{n-1}(\lambda) + A_{n-2}(\lambda) + 1}{\lambda^2} -1 \Bigr) \, . \nn
\end{align}
Applying Lemma \ref{AREC} in this identity yields 
\begin{align}
\alpha_n(\lambda^2) 
& \= \frac1{2\lambda^2} \Bigl( \lambda^2 (A_{n-1}(\lambda) - A_{n-2}(\lambda) )  + (\lambda^2-2w_{n-1})  \, (A_{n-1}(\lambda) + A_{n-2}(\lambda) + 1) - \lambda^2 \Bigr) \nn\\
& \= \frac1{\lambda^2} \Bigl( \lambda^2 A_{n-1}(\lambda) -  w_{n-1} \, (A_{n-1}(\lambda) + A_{n-2}(\lambda) + 1)  \Bigr) \,.\nn
\end{align}
The lemma is proved.
\epf 

Observe that $C_2(\lambda,\mu; n, {\bf s})$ satisfies the parity symmetries
$$
C_2(\lambda,\mu) \= C_2 (-\lambda,-\mu)\,, \qquad C_2(\lambda,-\mu)\= C_2(-\lambda,\mu)\,.
$$
So
\begin{align}
\sum_{i,j\geq 1} \frac{\p^2 \log \tau_n({\bf t})}{\p t_{2i-1}\p t_{2j-1}} \frac1{\lambda^{2i+1} \mu^{2j+1}} &  \= \frac{C_2(\lambda,\mu)-C_2(-\lambda,\mu)}2 \,.
\end{align}
\begin{lemma}\label{mainlemma} The following identity hold true:
\begin{align}
& \lambda\mu \frac{\alpha_n(\lambda^2)+\alpha_n(\mu^2)+2\alpha_n(\lambda^2)\alpha_n(\mu^2) - w_n\, w_{n-1}\, \bigl(\gamma_{n}(\lambda^2)\gamma_{n+2}(\mu^2) +\gamma_n(\mu^2)\gamma_{n+2}(\lambda^2)\bigr)}{(\lambda^2-\mu^2)^2} \nn\\
&  + \lambda\mu \frac{\alpha_{n+1}(\lambda^2)+\alpha_{n+1}(\mu^2)+2\alpha_{n+1}(\lambda^2)\alpha_{n+1}(\mu^2) - w_{n+1}\, w_n\, \bigl(\gamma_{n+1}(\lambda^2)\gamma_{n+3}(\mu^2) +\gamma_{n+1}(\mu^2)\gamma_{n+3}(\lambda^2)\bigr)}{(\lambda^2-\mu^2)^2} \nn\\
& = \; \frac{A_n(\lambda) + A_n(\mu)+2A_n(\lambda)A_n(\mu)-w_n\bigl(G_{n+1}(\lambda)G_n(\mu) + G_{n+1}(\mu)G_n (\lambda)\bigr)}{2(\lambda-\mu)^2}   \nn\\
& \quad - \frac{A_n(-\lambda) + A_n(\mu)+2A_n(-\lambda)A_n(\mu)-w_n\bigl(G_{n+1}(-\lambda)G_n(\mu) + G_{n+1}(\mu)G_n (-\lambda)\bigr)}{2(\lambda+\mu)^2}  \,. \label{toshow}
\end{align}
\end{lemma}
\pf 
Applying \eqref{g1}--\eqref{a1} and the parity symmetry $$A_n(-\lambda) \= A_n(\lambda)$$ we find that  it suffices to prove the following equality
\begin{align}
&   - \lambda\mu \+ 2 \, \lambda \mu \, \Bigl[A_{n-1}(\lambda) - \frac{w_{n-1}}{\lambda^2} \, \bigl(A_{n-1}(\lambda) + A_{n-2}(\lambda) + 1\bigr)  +\frac12\Bigr]   \nn\\
& \qquad\qquad\qquad\qquad \Bigl[A_{n-1}(\mu) - \frac{w_{n-1}}{\mu^2} \, \bigl(A_{n-1}(\mu) + A_{n-2}(\mu) + 1\bigr)  +\frac12\Bigr] \nn\\
& -  \frac{w_n\, w_{n-1}}{\lambda\mu} \, \Bigl[(A_{n-1}\bigl(\lambda) + A_{n-2}(\lambda) + 1\bigr)\bigl(A_{n+1}(\mu) + A_{n}(\mu) + 1\bigr) \nn\\
& \qquad \qquad \qquad + \bigl(A_{n-1}(\mu) + A_{n-2}(\mu) + 1\bigr)\bigl(A_{n+1}(\lambda) + A_{n}(\lambda) + 1\bigr)\Bigr] \nn\\
& -  \frac{w_{n+1}\, w_n}{\lambda\mu}\, \Bigl[ \bigl(A_{n}(\lambda) + A_{n-1}(\lambda) + 1\bigr)\bigl(A_{n+2}(\mu) + A_{n+1}(\mu) + 1\bigr) \nn\\
& \qquad \qquad \qquad  + \bigl(A_{n}(\mu) + A_{n-1}(\mu) + 1\bigr) \bigl( A_{n+2}(\lambda) + A_{n+1}(\lambda) + 1\bigr)\Bigr] \nn\\
&   +  2 \, \lambda  \mu \, \Bigl[A_{n}(\lambda) - \frac{w_{n}}{\lambda^2} \, \bigl(A_{n}(\lambda) + A_{n-1}(\lambda) + 1\bigr)  +\frac12\Bigr]   \nn\\
&\qquad\qquad\qquad\qquad  \Bigl[A_{n}(\mu) - \frac{w_{n}}{\mu^2} \, \bigl(A_{n}(\mu) + A_{n-1}(\mu) + 1\bigr)  +\frac12\Bigr ]   \nn\\
& \= \frac{(\lambda+\mu)^2}2 \Bigl[A_n(\lambda) + A_n(\mu)+2A_n(\lambda)A_n(\mu) \nn\\ 
& \qquad \qquad \qquad - \frac{w_n}{ \lambda  \mu}  \Bigl(
\bigl( A_{n+1}(\lambda) + A_{n}(\lambda) + 1 \bigr) \bigl(  A_{n}(\mu) + A_{n-1}(\mu) + 1\bigr) \nn\\
& \qquad\qquad\qquad +  \bigl( A_{n+1}(\mu) + A_{n}(\mu) + 1 \bigr) 
\bigl( A_{n}(\lambda) + A_{n-1}(\lambda) + 1\bigr)\Bigr)\Bigr]   \nn\\
& \qquad  - \frac{(\lambda-\mu)^2}2 \Bigl[ A_n(\lambda) + A_n(\mu)+2A_n(\lambda)A_n(\mu) \nn\\
& \qquad \qquad \qquad + \frac{w_n}{ \lambda  \mu}  \Bigl(
\bigl( A_{n+1}(\lambda) + A_{n}(\lambda) + 1 \bigr) \bigl(  A_{n}(\mu) + A_{n-1}(\mu) + 1\bigr) \nn\\
&\qquad\qquad\qquad +  \bigl( A_{n+1}(\mu) + A_{n}(\mu) + 1 \bigr) 
\bigl( A_{n}(\lambda) + A_{n-1}(\lambda) + 1\bigr)\Bigr)\Bigr]\,.  \nn
\end{align}
Noting that 
\begin{align}
& \lambda\mu\cdot {\rm lhs} \= \nn\\
& - \lambda^2\mu^2 \+ 2  \Bigl[\lambda^2 A_{n-1}(\lambda) - w_{n-1} \, \bigl(A_{n-1}(\lambda) + A_{n-2}(\lambda) + 1\bigr)  +\frac{\lambda^2}2\Bigr]  \nn\\
& \qquad\qquad\qquad\qquad \Bigl[\mu^2 A_{n-1}(\mu) - w_{n-1} \, \bigl(A_{n-1}(\mu) + A_{n-2}(\mu) + 1\bigr)  +\frac{\mu^2}2\Bigr] \nn\\
& -   w_{n-1} \, \Bigl[(A_{n-1}\bigl(\lambda) + A_{n-2}(\lambda) + 1\bigr) \Bigl( \mu^2 \bigl(A_{n}(\mu)-A_{n-1}(\mu)\bigr) +w_{n-1}\bigl(A_{n-1}(\mu) + A_{n-2}(\mu) + 1\bigr) \Bigr) \nn\\
&\qquad\quad + \bigl(A_{n-1}(\mu) + A_{n-2}(\mu) + 1\bigr) \Bigl( \lambda^2 \bigl(A_{n}(\lambda)-A_{n-1}(\lambda)\bigr) +w_{n-1}\bigl(A_{n-1}(\lambda) + A_{n-2}(\lambda) + 1\bigr) \Bigr)\Bigr] \nn\\
& - w_n \, \Bigl[ \bigl(A_{n}(\lambda) + A_{n-1}(\lambda) + 1\bigr) \Bigl( \mu^2 \bigl(A_{n+1}(\mu)-A_n(\mu)\bigr) +w_n\bigl(A_{n}(\mu) + A_{n-1}(\mu) + 1\bigr) \Bigr) \nn\\
& \qquad\qquad + \bigl(A_{n}(\mu) + A_{n-1}(\mu) + 1\bigr) \Bigl( \lambda^2 \bigl(A_{n+1}(\lambda)-A_n(\lambda)\bigr) +w_n\bigl(A_{n}(\lambda) + A_{n-1}(\lambda) + 1\bigr) \Bigr)\Bigr] \nn\\
& + 2 \Bigl[ \lambda^2 A_{n}(\lambda) - w_{n} \bigl(A_{n}(\lambda) + A_{n-1}(\lambda) + 1\bigr)  +\frac{\lambda^2}2\Bigr]  
 \Bigl[ \mu^2 A_{n}(\mu) - w_{n} \bigl(A_{n}(\mu) + A_{n-1}(\mu) + 1\bigr)  +\frac{\mu^2}2\Bigr ]   \nn
\end{align}
and that
\begin{align}
\lambda\mu\cdot {\rm rhs} \= & 2 \, \lambda^2\mu^2   \Bigl(A_n(\lambda) + A_n(\mu)+2A_n(\lambda)A_n(\mu)\Bigr) \nn\\ 
& + w_n \, ( \lambda^2+\mu^2 ) \Bigl(
\bigl( A_{n+1}(\lambda) + A_{n}(\lambda) + 1 \bigr) \bigl(  A_{n}(\mu) + A_{n-1}(\mu) + 1\bigr)  \nn\\
& \qquad\qquad\qquad\qquad +  \bigl( A_{n+1}(\mu) + A_{n}(\mu) + 1 \bigr) \bigl( A_{n}(\lambda) + A_{n-1}(\lambda) + 1\bigr)\Bigr) \,,  \nn
\end{align}
we find  
\begin{align}
&\lambda\mu\cdot ({\rm lhs} -{\rm rhs})  \nn\\
 \= &  \lambda^2\mu^2 (2 A_{n-1}(\lambda)  A_{n-1}(\mu)-2 A_{n}(\lambda) A_{n}(\mu) + A_{n-1}(\lambda)+A_{n-1}(\mu) -A_{n}(\lambda)-A_{n}(\mu)\bigr)\nn\\
 & -\mu^2 (A_{n-1}(\mu)+A_{n}(\mu)+1) \bigl(w_{n-1} (A_{n-2}(\lambda)+A_{n-1}(\lambda)+1)-w_n (A_{n}(\lambda)+A_{n+1}(\lambda)+1)\bigr) \nn\\
 & - \lambda^2 \bigl(A_{n-1}(\lambda)+A_{n}(\lambda)+1) \bigl(w_{n-1} (A_{n-2}(\mu)+A_{n-1}(\mu)+1)-w_n (A_{n}(\mu)+A_{n+1}(\mu)+1)\bigr) \nn \\
 \= &  \lambda^2\mu^2 (2 A_{n-1}(\lambda)  A_{n-1}(\mu)-2 A_{n}(\lambda) A_{n}(\mu) + A_{n-1}(\lambda)+A_{n-1}(\mu) -A_{n}(\lambda)-A_{n}(\mu)\bigr)\nn\\
 & +\lambda^2 \mu^2  (A_{n-1}(\mu)+A_{n}(\mu)+1) (A_n(\lambda)-A_{n-1}(\lambda)) \nn\\
 & + \lambda^2 \mu^2 (A_{n-1}(\lambda)+A_{n}(\lambda)+1) (A_n(\mu)-A_{n-1}(\mu)) \= 0 \nn \,,
\end{align}
where Lemma~\ref{AREC} is used. The lemma is proved. \epf

\noindent {\it End of proof of Theorem \ref{maintheorem}}.  It follows from Lemma \ref{mainlemma} that 
$$
\frac{\p^2 \log \tau_n({\bf s})}{\p s_i \p s_j} \= \frac{\p^2}{\p s_i \p s_j} (\Lambda +1) \, \log \tau_n^{\textsc{\tiny\rm dKdV}}({\bf s}) \,. 
$$
Combining with \eqref{diff} we find that 
$$
f({\bf s}) \= \beta_0  + \sum_{k\geq 1} \beta_k s_k \,, 
$$ 
where $\beta_0,\beta_1,\beta_2,\cdots$ are constants (independent of~$n$). The theorem is proved. \epf

\section{Proofs of Theorem~\ref{thm3} and Corollary~\ref{app_hodge}} \label{appsection}

In this section, using 
Proposition~\ref{main1}, Corollary~\ref{cor1} and Theorem~\ref{maintheorem}, 
we are going to prove Theorem~\ref{thm3} and Corollary~\ref{app_hodge}.

\subsection{Ribbon graphs with even valencies} 
In this subsection, we first prove Theorem~\ref{thm3}, then we give a further study to the modified GUE partition function with even couplings. 

\pf of Theorem~\ref{thm3}.  \quad 
Define  $\F_n({\bf s})$ and $Z_n({\bf s})$ by
\begin{align}
& \F_n({\bf s}) \:= \frac{n^2}{2} \Bigl( \log n -\frac32\Bigr) \,-\, \frac1{12} \log n \+ \sum_{g\geq 2}  \frac{B_{2g}}{4g(g-1) \, n^{2g-2}} \nn\\
& \qquad\qquad\qquad\quad \+ \sum_{k\geq 0} \frac1{k!} \sum_{j_1, \dots, j_k\geq 1} \bll\tr \, M^{2j_1} \, \cdots \, \tr \, M^{2j_k} \brr_c \, s_{j_1} \cdots s_{j_k} \, ,  \nn\\
& Z_n({\bf s}) \:= e^{\F_n({\bf s})} \, . \label{FZzeta}
\end{align}
Here $B_m$ denotes the $m^{\rm th}$ Bernoulli number. 
Then $Z_n({\bf s})$ is a particular tau-function (of the discrete KdV hierarchy) reduced from the Toda lattice hierarchy \cite{DuY1}. The initial value of 
$w_n({\bf s}):= \frac{Z_{n+1}({\bf s}) \, Z_{n-1}({\bf s})} {Z_n({\bf s})^2}$ is given by $w_n({\bf s}={\bf 0})= n$. 
The theorem then follows from Lemma \ref{usefullemma}, Corollary \ref{cor1}, as well as the Theorem 1.1.1 of \cite{DuY1}.  
\epf

Define a formal series $Z(x,{\bf s};\epsilon)$ by
\begin{align} 
\log Z(x, {\bf s};\epsilon) \:=  & \, \frac{x^2}{2\epsilon^2} \Bigl( \log x -\frac32\Bigr) -\frac1{12} \log x 
 \+ \sum_{g\geq 2} \epsilon^{2g-2} \frac{B_{2g}}{4g(g-1)x^{2g-2}} \nn\\
& \+ \sum_{g\geq 0} \epsilon^{2g-2}  \sum_{k\geq 1} \sum_{j_1, \dots, j_k\geq 1 \atop |j|\geq 2g-2+k} a_g(2j_1, \dots, 2j_k) \, s_{j_1} \cdots s_{j_k} \, x^{2-2g -k+|j|} \, . \label{taugue}
\end{align} 
Here, $x$ is the t'Hooft coupling constant~\cite{thooft1,thooft2}. 
Recall that we could view $Z(x, {\bf s}; \epsilon)$ as 
 a tau-function {\it reduced from the Toda lattice} of the 
discrete KdV hierarchy under the identification $n=x/\epsilon$ as well as the flow rescalings 
$\p_{s_j}\mapsto \e \, \p_{s_j}$. More precisely,
 define $$w(x,{\bf s};\e) \:= \frac{Z(x+\epsilon,{\bf s}; \e) \,  Z(x-\epsilon,{\bf s}; \e)}{Z(x,{\bf s};\e)^2} \,, $$
 then $w(x,{\bf s};\e)$ is a particular solution to the discrete KdV hierarchy:
 $$
 \e \, \frac{\p L}{\p s_j} \= \bigl[ A_{2j-1} \,,\, L \bigr] 
 $$
 with
 $L := \Lambda^2 \+ w(x+\e) + w(x) \+ w(x) \, w(x-\e)  \, \Lambda^{-2}$, $A_{2j-1}:= L^j$, $\Lambda:=e^{\e \p_x}$. 
 Validity of these statements can be found in the 
 Appendix of~\cite{DuY1}. The initial data of this solution is given by
 \beq\label{iniw}
 w(x, {\bf 0}; \epsilon) \; \equiv \; x \= n \, \epsilon \,. 
 \eeq
 Let $Z^{\textsc{\tiny\rm dKdV}}(x,{\bf s};\e)$ be the tau-function of the solution $w(x,{\bf s};\e)$. 
 The following corollary follows from Theorem~\ref{maintheorem}. 
 \begin{cor} \label{cortheorem1} There exist constants $\alpha,\beta_0,\beta_1,\beta_2,\cdots$ such that 
\beq 
Z(x, {\bf s};\e) \= e^{\alpha \, x + \beta_0  + \sum_{k\geq 1} \beta_j s_j} \, 
Z^{\textsc{\tiny\rm dKdV}}(x,{\bf s};\e) \, Z^{\textsc{\tiny\rm dKdV}}(x+\e, {\bf s};\e) \,.
\eeq
 \end{cor}
Note that the constants $\alpha,\beta_0,\beta_1,\beta_2,\cdots$ right above now {\it can} depend on~$\e$.  
In what follows, we fix the ambiguities simply by requiring 
$Z^{\textsc{\tiny\rm dKdV}}(x,{\bf s};\e)$
to be the unique function satisfying
\beq\label{definitionzdkdv}
Z(x, {\bf s};\e) \=  \, Z^{\textsc{\tiny\rm dKdV}}(x,{\bf s};\e) \, Z^{\textsc{\tiny\rm dKdV}}(x+\e, {\bf s};\e)\,. 
\eeq

\begin{remark}
The following formal series of~${\bf s}$ 
\beq\label{modZ}
Z^{\textsc{\tiny\rm dKdV}}\Bigl(x+\frac\e2, {\bf s};\e\Bigr)=:\widetilde Z(x,{\bf s};\e)
\eeq
was introduced in~\cite{DLYZ2} by Si-Qi Liu, Youjin Zhang and the authors of the present paper, 
called the {\it modified GUE partition function with even couplings}, 
which plays an important role in a proof of the Hodge--GUE correspondence~\cite{DLYZ2}. 
Moreover, Liu, Zhang and the authors derived the  
{\it Dubrovin--Zhang loop equation} for~$\log\widetilde Z$
 from the corresponding Virasoro constraints, 
which also provides an algorithm for computing the modified GUE
correlators of an arbitrary genus~\cite{DLYZ2}. 
Very recently, Jian Zhou~\cite{Zhou2} derived the {\it topological recursion of 
Chekhov--Eynard--Orantin type} for the modified GUE correlators from  
the Virasoro constraints constructed in~\cite{DLYZ2}; moreover, as a 
consequence of the topological recursion, an interesting formula between  
intersection numbers and $k$-point functions of modified GUE correlators 
was obtained by Zhou~\cite{Zhou2} (see the Theorem~3 in~\cite{Zhou2} for the details); 
it remains an open question to match the formula of~Zhou with  
another interesting formula obtained by Ga\"etan Borot and 
Elba Garcia-Failde~\cite{BGF} (see the Corollary~12.3 of~\cite{BGF}) 
as a consequence of the Hodge--GUE correspondence
(or with a slightly different but equivalent consequence like~\eqref{otherconsequence2} in below), which may lead to a new 
proof of the Hodge--GUE correspondence. 
Last but not least, as a corollary of Theorem~\ref{thm3}, let us 
 give a third algorithm of computing the modified GUE correlators 
 based on the following {\it full genera} formulae:
\begin{align}
& \e^2 \sum_{j_1,j_2\geq 0} \frac{\langle \phi_{j_1} \phi_{j_2}\rangle(x;\e)}{\lambda^{j_1+1}_1 \lambda^{j_2+1}_{2}}  
\= \frac{\tr\, \bigl[R_{\frac{x}\e+\frac12}(\frac{\lambda_1}\e) R_{\frac{x}\e+\frac12}(\frac{\lambda_2}\e)\bigr]}{(\lambda_1-\lambda_2)^2} -\frac1{(\lambda_1-\lambda_2)^2} \,, \label{twopointmod} \\
& \e^k \sum_{j_1,\dots,j_k\geq 0} 
\frac{\langle \phi_{j_1}\cdots \phi_{j_k}\rangle(x;\e)}{\lambda^{j_1+1}_1\cdots \lambda^{j_k+1}_{k}}  \= 
-\frac1{k}\sum_{\sigma\in S_k} \frac{ \tr \,\bigl[ R_{\frac{x}\e+\frac12}(\frac{\lambda_{\sigma_1}}{\e}) \cdots 
R_{\frac{x}\e+\frac12}(\frac{\lambda_{\sigma_k}}\e)\bigr]}
{\prod_{\ell=1}^k (\lambda_{\sigma_\ell}-\lambda_{\sigma_\ell+1})}   \quad (k\geq 3)\,,   \label{kpointmod}
\end{align}
where $\langle \phi_{j_1}\cdots \phi_{j_k}\rangle(x;\e)$ denote the modified GUE 
correlators with even couplings, defined by
\beq\label{modifiedcorrelators}
\langle \phi_{j_1}\cdots \phi_{j_k}\rangle(x;\e) \:= 
\frac{\p^k\log \widetilde Z}{\p s_{j_1}\dots \p s_{j_k}}(x,{\bf s}={\bf 0};\e) \, ,
\eeq
and 
$R_n(\lambda)$ is defined in Definition~\ref{defRAB}. 
We notice that the reason that one can talk about ``genus" for $\log \widetilde Z$ is because 
$\log \widetilde Z$ is even in~$\e$ and so are $\langle \phi_{j_1}\cdots \phi_{j_k}\rangle(x;\e)$. 
A concrete algorithm using the formulae of 
the form \eqref{twopointmod}--\eqref{kpointmod} for 
calculating the corresponding correlators including certain large genus asymptotics 
is given in~\cite{DuY3}. 
\end{remark}

\begin{remark}
Very recently it was shown in~\cite{GGR} 
that $Z^{\textsc{\tiny\rm dKdV}}(x,{\bf s};\e)$ and $Z^{\textsc{\tiny\rm dKdV}}(x+\e,{\bf s};\e)$ are identified with 
the LUE partition functions with $\alpha=-1/2$ and $\alpha=1/2$, respectively.
One can obtain their $k$-point series by putting $x\rightarrow x\mp\frac\e2$ in~\eqref{twopointmod}--\eqref{kpointmod}. 
An interesting genus expansion for $Z^{\textsc{\tiny\rm dKdV}}(x,{\bf s};\e)$ was discovered in~\cite{CDOC}. The interplay between 
$Z^{\textsc{\tiny\rm dKdV}}$ and $\widetilde Z$ suggests a Hurwitz/Hodge correspondence that deserves a further study.
\end{remark}


Using the definitions of~$Z(x, {\bf s};\e)$ and~$\widetilde{Z} (x,{\bf s};\e)$ and using the expansion
$$
\frac2{e^z+ e^{-z}} \;=:\;  \sum_{k\geq 0} \frac{E_k}{k!} z^k \,, 
$$
with $E_k$, $k\geq 0$ being the Euler numbers, we have the following formula:
\begin{align}
& \log \widetilde{Z} (x,{\bf s};\e) \nn\\
& = \,  \Bigl(\frac14 \log x - \frac38 \Bigr) \frac{x^2}{\e^2} \,-\, \frac{5}{48} \log x  \+ \sum_{g\geq 2} \frac{\e^{2g-2}}{4g(2g-1)(2g-2) x^{2g-2}}\sum_{g'=0}^g (2g'-1)\binom{2g}{2g'} \frac{E_{2g-2g'} B_{2g'}}{2^{2g-2g'}} \nn\\
& \quad \+ \sum_{h\geq 0} \epsilon^{2h-2}  \sum_{g,r\geq 0 \atop g+r=h} \sum_{k\geq 1} 
\sum_{j_1, \dots, j_k\geq 1 
} 
\binom{2-2g -k+|j|}{2r} \frac{E_{2r}}{2^{2r}} 
a_g(2j_1, \dots, 2j_k) \, s_{j_1} \cdots s_{j_k} \, x^{2-2h -k+|j|} \,. \nn
\end{align}
In other words, the modified GUE correlators with even couplings~\eqref{modifiedcorrelators} have the expressions:
\beq
\langle \phi_{j_1}\cdots \phi_{j_k}\rangle(x;\e) \= k!
\sum_{h\geq 0} \epsilon^{2h-2} x^{2-2h -k+|j|} \sum_{g,r\geq 0 \atop g+r=h} 
 \binom{2-2g -k+|j|}{2r} \frac{E_{2r}}{2^{2r}} 
a_g(2j_1, \dots, 2j_k) \,, \label{modifiedcorranda}
\eeq
where $k\geq 1$ and $j_1,\dots,j_k\geq 1$.  It should be noted that the $\langle \phi_{j_1}\cdots \phi_{j_k}\rangle(x;\e)$ with 
$k\geq 1$, $j_1,\dots,j_k\geq 1$ is a polynomial of~$x$. 

\subsection{Combinations of certain special cubic Hodge integrals}  \label{section52}
Based on the Hodge--GUE correspondence and using Theorem~\ref{thm3}, we   
compute in this subsection combinations of certain special cubic Hodge integrals. More precisely, 
we will prove Corollary~\ref{app_hodge}.

The cubic Hodge free energy associated with 
$\Lambda_g(-1)\,\Lambda_g(-1)\,\Lambda_g\bigl(\tfrac12\bigr)$ is defined by
$$
\mathcal H(\bt;\epsilon) \= \sum_{g\geq 0} \epsilon^{2g-2} \sum_{k\geq 0}  \frac 1{k!} 
\sum_{i_1, \dots, i_k\geq 0} t_{i_1}\cdots t_{i_k}\int_{\overline{\mathcal M}_{g,k}} 
\Lambda_g(-1)\,\Lambda_g(-1)\,\Lambda_g\bigl(\tfrac12\bigr)\, \psi_1^{i_1}\cdots \psi_k^{i_k} \, .
$$
Here, $\bt=(t_0,t_1,\dots)$. (Warning: Avoid from confusing with the variables 
$t_\ell$, $\ell\geq 0$ of the Toda lattice hierarchy used in Section~\ref{SectionMaintheorem}.) 
The {\it Hodge--GUE correspondence} connects $\mathcal H(\bt;\epsilon)$ with the GUE partition function 
with even couplings, which is given by the following theorem.

\noindent {\bf Theorem A.} (\cite{DLYZ2,DuY2})  The following identity holds true:
\begin{align}
&  \log Z (x, {\bf s};\epsilon) \+ 
\epsilon^{-2} \biggl(-\frac12  \sum_{j_1,j_2\geq 1} \frac{j_1\, j_2}{j_1+j_2}  \,  \bar{s}_{j_1}  \, \bar{s}_{j_2} 
+ \sum_{j\geq 1} \frac{j}{1+j}\, \bar{s}_{j} - x \,  \sum_{j\geq 1} \,  \bar{s}_{j}   - \frac14  + x \biggr)  \nn\\
&  \qquad\qquad\qquad \=   \mathcal{H} \Bigl(\bt\bigl(x-\frac\epsilon2, {\bf s}\bigr); \sqrt{2} \epsilon \Bigr) \+  
\mathcal{H} \Bigl(\bt\bigl(x+\frac\epsilon2, {\bf s}\bigr); \sqrt{2} \epsilon \Bigr)  \,, \label{hodgegue}
\end{align}
where $\bar{s}_{j} := \binom{2j}{j} \, s_j$ and 
\beq\label{timesubstitute}
t_i(x, {\bf s}) \:= \sum_{j\geq 1} j^{i+1} \bar{s}_j  \,-\, 1 \+ \delta_{i,1} \+ x \, \delta_{i,0} \,, \qquad i\geq 0\,.
\eeq

Recall from~\eqref{modZ} that the modified GUE partition function with even couplings $\widetilde Z$ is defined as the 
unique series of $x-1$ and ${\bf s}$ satisfying 
\beq
Z(x, {\bf s};\e) \=  \, \widetilde Z \Bigl(x-\frac\e2,{\bf s};\e\Bigr) \, \widetilde Z \Bigl(x+\frac\e2, {\bf s};\e\Bigr)\,. \label{ZZZ}
\eeq
Combining~\eqref{ZZZ} with~\eqref{hodgegue} we obtain the following corollary.
\begin{cor} \label{matchingdtauandh} The following formula holds true:
\begin{align}
& \log \widetilde Z \bigl(x,{\bf s}; \e\bigr) \= \cH\bigl(\bt \bigl(x, {\bf s}\bigr); \sqrt{2} \e \bigr)  \+ \frac1{4 \e^2}  \sum_{j_1,j_2\geq 1} \frac{j_1\,j_2}{j_1+j_2}  \,  
\bar{s}_{j_1}  \, \bar{s}_{j_2}    \+ \frac{x}{2 \e^2}\Bigl(\sum_{j\geq 1} \bar s_j-1\Bigr) \nn\\
& \qquad \qquad \qquad   \qquad\qquad  - \frac1 {2 \e^2} \sum_{j\geq 1} \frac{j}{1+j}\, \bar{s}_{j} 
 \+ \frac1{8\e^2} \, . \label{dkdvtauhodge}
\end{align}
\end{cor}

Denote $ \Omega_{g}:=\Lambda_{g}(-1)\,\Lambda_{g}(-1)\,\Lambda_{g}\bigl(\frac12\bigr)$ as in the introduction, and write 
 $$\Omega_{g}=: \sum_{d\geq 0} \Omega_{g}^{[d]}\,, \qquad \Omega_{g}^{[d]} \in H^{2d}(\overline{\mathcal{M}}_{g,k})\,.$$
It might be helpful to notice that for $g=1$, $\deg \, \Omega_{1} \leq 1$;  for $g\geq 2$, $\deg \, \Omega_{g} \leq 3g-3$. 
Motivated by Theorem~A, let us consider the following combination of Hodge integrals.
For any given $k\geq0$, $i_1,\dots,i_k\geq 0$, define a formal series $H_{i_1,\dots,i_k}(x; \e)\in \e^{-2} \QQ[[x-1,\e^2]]$ by
\beq
H_{i_1,\dots,i_k}(x; \e) \:=  2^{g-1} 
\sum_{g=0}^\infty \epsilon^{2g-2}   \sum_{d= 0}^{3g} \sum_{\lambda\in \mathbb{Y}} \frac{(-1)^{\ell(\lambda)}}{m(\lambda)! } 
\bll \Omega_g^{[d]} \, e^{(x-1)\tau_0}  \,  \tau_{\lambda+1} \, \tau_{i_1} \cdots \tau_{i_k}  \brr_g \,.
\eeq
\noindent Note that in the notation $\langle \,\dots\, \rangle_{g}$\,, we omit the index $m$ from $\langle \,\dots\, \rangle_{g,m}$. 
For such an abbreviation, $m$ should be recovered from counting the number of $\tau$'s in $``\dots"$. Therefore, for each fixed $g,d$ and for each monomial in the 
Taylor expansion $e^{(x-1)\tau_0}= \sum_{r=0}^\infty \frac1{r!} (x-1)^r$, the above summation over partitions $\sum_{\lambda\in \mathbb{Y}}$ 
is a finite sum, i.e., the degree-dimension matching $|\lambda| = 3g-3+k+r-d-|i|$ has to be hold. Lemma~\ref{van1} also easily follows from this constrain with $r=0$ taken. 
The numbers $H_{g,i_1,\dots,i_k}$ defined by~\eqref{defhgi} and the formal series $H_{i_1,\dots,i_k}(x; \e)$ are clearly related by
\beq
H_{i_1,\dots,i_k}(x=1; \e) \= \sum_{g=0}^\infty \e^{2g-2} H_{g,i_1,\dots,i_k}. 
\eeq
\begin{prop} For any $k\geq 0$ and $j_1,\dots,j_k\geq 1$, the following formula holds true:
\begin{align}
\phi_{j_1,\dots,j_k}(x;\e)  \; = \;  & \prod_{\ell=1}^k \binom{2j_\ell}{j_\ell} \,  
\sum_{i_1,\dots,i_k\geq 0}   \prod_{\ell=1}^k j_\ell^{i_\ell+1} \, H_{i_1,\dots,i_k}\bigl(x; \e\bigr)  \nn\\
& + \, \frac{\delta_{k,2}}{2\e^2} \frac{j_1 \, j_2}{j_1+j_2} \binom{2j_1}{j_1} \binom{2j_2}{j_2} - \frac{\delta_{k,1}}{2\e^2}  \binom{2j_1}{j_1} 
\biggl( \frac{j_1}{1+j_1} - x \biggr) \,.  \label{otherconsequence2}
\end{align}
\end{prop}
\pf
Note that the $\cH\bigl(\bt ; \sqrt{2} \e \bigr)$ has the expression
\[
\cH\bigl(\bt ; \sqrt{2} \e \bigr) \= 
\sum_{g} 2^{g-1} \e^{2g-2}  
\sum_{m_0,m_1,m_2,\cdots} 
\int_{\overline{\mathcal M}_{g,\sum_i {m_i}}} 
 \Omega_{g}
\, \prod_{s=1}^{m_0} \psi_s^{0}  \prod_{s=m_0+1}^{m_0+m_1} \psi_{s}^1 \prod_{s=m_0+m_1+1}^{m_0+m_1+m_2} \psi_{s}^2 \cdots 
\prod_{i=0}^\infty \frac{t_i^{m_i}} {m_i!}  \,.
\]
Formula~\eqref{otherconsequence2} is then proved by substituting~\eqref{timesubstitute} and by 
using Corollary~\ref{matchingdtauandh}. 
\epf


\pf of Corollary~\ref{app_hodge}. Note that the $k=0$ case is already given in~\cite{DLYZ2}.
By taking $x=1$ in~\eqref{otherconsequence2} and using Theorem~\ref{thm3} we find~\eqref{kgeq2}.
Formula~\eqref{kequals1} is then implied in a standard way by using the following linear equation (proven in~\cite{DLYZ2})
\beq\label{stringtype}
\sum_{k\geq 1} k s_k \frac{\p \widetilde Z}{\p s_k} \+ \Bigl(\frac{x^2}{4\e^2}-\frac1{16} \Bigr) \widetilde Z \= \frac12 \frac{\p \widetilde Z}{\p s_1}
\eeq
and the fact that 
\beq 
a_{n,j} \= \frac{\p^2 \log \tau_n^{\textsc{\tiny\rm dKdV}}}{\p s_1 \p s_j} \,,\qquad j\geq 1\,.\label{residuedeftau}
\eeq
Here $\widetilde Z=\widetilde Z(x,{\bf s};\e)$ denotes the modified GUE partition function with even couplings. 
Note that the fact~\eqref{residuedeftau} can be obtained by taking the coefficients of~$\lambda^{-1}$ 
on the both sides of~\eqref{d1}. The corollary is proved.
\epf

\begin{appendix}
\section{On consequence of the Hodge--GUE correspondence}\label{appendixa}
In this appendix, we derive a consequence of the Hodge--GUE correspondence that 
has a similar flavour to formula~\eqref{otherconsequence2}. Note that 
\begin{align}
& \mathcal{H}\bigl(\bt(x,{\bf s}); \sqrt{2}\e\bigr) \nn\\
& ~ \= \sum_{g\geq 0} 2^{g-1} \e^{2g-2} \sum_{k\geq 0}  \frac1{k!}
\int_{\overline{\mathcal{M} }_{g,k}}  \Omega_{g,k}   
\prod_{m=1}^k \biggl(\sum_{i_m\geq 0} t_{i_m}(x,{\bf s}) \psi_m^{i_m}\biggr)\nn \\
& ~ \= \sum_{g\geq 0} 2^{g-1} \e^{2g-2} \sum_{k\geq 0}  \frac1{k!}
\int_{\overline{\mathcal{M} }_{g,k}}  \Omega_{g,k}  \binom{k}{l}  
\prod_{m=l+1}^k  \biggl((x -1) -  \frac{\psi_m^2}{1-\psi_m} \biggr)  
\sum_{ p_1,\dots,p_l}   \prod_{m=1}^{l}\frac{p_m\bar{s}_{p_m}}{1- p_m\psi_m} \,.   \nn
\end{align}
Then by comparing the coefficients of $s_{p_1} \dots s_{p_l}$ of the both sides of~\eqref{dkdvtauhodge} we get
\begin{align}
& \langle\sigma_{p_1}\dots \sigma_{p_l}\rangle_g(x) \= 
  \sum_{k\geq l}  \frac{1}{(k-l)!}
\int_{\overline{\mathcal{M} }_{g,k}}  \Omega_{g,k}   
\prod_{m={l+1}}^k  \biggl((x -1) -  \frac{\psi_m^2}{1-\psi_m} \biggr)  
  \prod_{m=1}^{l}\frac{p_m \binom{2p_m}{p_m}}{1- p_m\psi_{m}} \nn\\
  & \qquad\qquad\qquad\qquad \+ \delta_{g,0} \delta_{l,2} \, \frac{p_1 \, p_2}{p_1+p_2}  
  \binom{2p_1}{p_1} \binom{2p_2}{p_2} \+  \frac12\delta_{g,0} \delta_{l,1} \binom{2p_1}{p_2} \biggl( \frac{p_1}{1+p_1} - x \biggr) \,. \label{appeq1}
\end{align}
Here $\langle\sigma_{p_1}\dots \sigma_{p_l}\rangle(x;\e) =: \sum_{g\geq 0} \e^{2g-2} \langle\sigma_{p_1}\dots \sigma_{p_l}\rangle_g(x)$, 
and $\langle\sigma_{p_1}\dots \sigma_{p_l}\rangle(x;\e) $ are the 
modified GUE correlators with even couplings defined in~\eqref{modifiedcorrelators}. Taking $x=1$ we find
\begin{align}
\langle\sigma_{p_1}\dots \sigma_{p_l}\rangle_g|_{x=1} \= 
 \sum_{k\geq l}  \frac{1}{(k-l)!}
\int_{\overline{\mathcal{M} }_{g,k}}  \Omega_{g,k}   
\prod_{m={l+1}}^k  \biggl(-  \frac{\psi_m^2}{1-\psi_m} \biggr)  
  \prod_{m=1}^{l}\frac{p_m \binom{2p_m}{p_m}}{1- p_m\psi_{m}} \,. \label{123}
\end{align}
A further consideration to~\eqref{123} was given in~\cite{BGF}.

Combining~\eqref{appeq1} with~\eqref{modifiedcorranda} we find for any fixed $l\geq1$, $p_1,\dots,p_l\ge1$ the following identities: 
\begin{align}
& l! \, x^{2-2g -l+|j|} \sum_{g_1,r\geq 0 \atop g_1+r=g} 
 \binom{2-2g_1-l+|j|}{2r} \, \frac{E_{2r}}{2^{2r}} 
\,a_{g_1}(2p_1, \dots, 2p_l) \nn\\
& \quad \= 
\sum_{k\geq l}  \frac{1}{(k-l)!}
\int_{\overline{\mathcal{M} }_{g,k}}  \Omega_{g,k}   
\prod_{m={l+1}}^k  \biggl((x -1) -  \frac{\psi_m^2}{1-\psi_m} \biggr)  
  \prod_{m=1}^{l}\frac{p_m \binom{2p_m}{p_m}}{1- p_m\psi_{m}} \nn\\
& \qquad\quad \+ \delta_{g,0} \delta_{l,2} \, \frac{p_1 \, p_2}{p_1+p_2}  
  \binom{2p_1}{p_1} \binom{2p_2}{p_2} \+  \frac12\delta_{g,0} \delta_{l,1} \binom{2p_1}{p_2} \biggl( \frac{p_1}{1+p_1} - x \biggr)\,, \qquad g\geq0\,.\label{relations}
\end{align}
Note that for any $g\geq0$, the RHS is a~priori a power series of~$x-1$, but the LHS shows that it is actually a monomial of~$x$ and so is 
also a polynomial of $x-1$. This subset of the identities deserve a further investigation. Moreover, 
the LHS vanishes when $g$ is sufficiently large, an so is the RHS; this provides another subset of the identities for the  
cubic Hodge integrals.


\end{appendix}

\medskip
\medskip
\medskip

\noindent Boris Dubrovin

\noindent SISSA, via Bonomea 265, Trieste 34136, Italy

\noindent dubrovin@sissa.it

\medskip

\noindent Di Yang

\noindent School of Mathematical Sciences, USTC, Hefei 230026, P.R.~China

\noindent diyang@ustc.edu.cn

\end{document}